\documentclass[a4paper,10pt]{article}
\pdfoutput=1
\usepackage{amsmath}    
\usepackage{amssymb}             
\usepackage{todonotes}
\usepackage{color,soul}
\usepackage[ngerman,british]{babel}
\usepackage{graphicx}
\usepackage{cite}
\usepackage{slashed}
\usepackage{xspace}
\usepackage{hyperref}
\usepackage[utf8]{inputenc}

\evensidemargin \oddsidemargin
\newcommand{\xf}{\hat X_t}

\newcommand{\tb}{\tan\beta}
\newcommand{\sbe}{s_\beta}
\newcommand{\cbe}{c_\beta}
\newcommand{\tbe}{t_\beta}
\newcommand{\sbb}{s_\beta^2}

\newcommand{\DR}{{\overline{\text{DR}}}}
\newcommand{\MS}{{\overline{\text{MS}}}}
\newcommand{\OS}{{\text{OS}}}
\newcommand{\EFT}{{\text{EFT}}}
\newcommand{\SM}{{\text{SM}}}

\newcommand{\MSSM}{{\text{MSSM}}}
\newcommand{\nonSM}{{\text{nonSM}}}
\newcommand{\SUSY}{{\text{SUSY}}}

\newcommand{\FH}{\mbox{{\tt Feyn\-Higgs}}\xspace}
\newcommand{\SHD}{\mbox{{\tt SUSY\-HD}}\xspace}
\newcommand{\FS}{\mbox{{\tt Flexible\-SUSY}}\xspace}
\newcommand{\FEFT}{\mbox{{\tt Flexible\-EFTHiggs}}\xspace}
\newcommand{\Fig}[1]{Fig.~\ref{#1}}
\newcommand{\Sec}[1]{Section~\ref{#1}}
\newcommand{\App}[1]{App.~\ref{#1}}
\newcommand{\Eq}[1]{Eq.~(\ref{#1})}
\newcommand{\Eqs}[2]{Eqs.~(\ref{#1}) and (\ref{#2})}
\newcommand{\Eqss}[2]{Eqs.~(\ref{#1})-(\ref{#2})}

\newcommand{\MZ}{M_Z}
\newcommand{\MW}{M_W}
\newcommand{\Mh}{M_h}

\newcommand{\MA}{M_A}

\newcommand{\Mstope}{M_{\tilde t_1}}
\newcommand{\Mstopz}{M_{\tilde t_2}}
\newcommand{\mstope}{m_{\tilde t_1}}
\newcommand{\mstopz}{m_{\tilde t_2}}
\newcommand{\SW}{s_\mathrm{w}}
\newcommand{\CW}{c_\mathrm{w}}

\newcommand{\cp}{\ensuremath{{\cal CP}}}

\newcommand{\msusy}{M_\SUSY}
\newcommand{\Xt}{X_t}

\newcommand{\tev}{\,\, \mathrm{TeV}}
\newcommand{\gev}{\,\, \mathrm{GeV}}
\newcommand{\mev}{\,\, \mathrm{MeV}}

\newcommand{\order}[1]{\ensuremath{{\cal O}(#1)}}
\newcommand{\al}{\alpha}
\newcommand{\als}{\al_s}
\newcommand{\alt}{\al_t}
\newcommand{\alb}{\al_b}

\begin{document}

\thispagestyle{empty}
\def\thefootnote{\fnsymbol{footnote}}

\begin{flushright}
DESY 17-072\\
IFT-UAM/CSIC-17-047\\
MPP-2017-108
\end{flushright}
\vspace{3em}
\begin{center}
{\Large\bf Reconciling EFT and hybrid calculations\\[.5em]
of the light MSSM Higgs-boson mass}
\\
\vspace{3em}
{
Henning Bahl$^a$\footnote{email: hbahl@mpp.mpg.de},
Sven Heinemeyer$^{b,c,d}$\footnote{email: sven.heinemeyer@cern.ch},
Wolfgang Hollik$^a$\footnote{email: hollik@mpp.mpg.de},
Georg Weiglein$^e$\footnote{email: georg.weiglein@desy.de}
}\\[2em]
{\sl ${}^a$Max-Planck Institut f\"ur Physik, F\"ohringer Ring 6, D-80805 M\"unchen, Germany}\\
{\sl ${}^b$Campus of International Excellence UAM+CSIC, Cantoblanco, E-28049 Madrid, Spain}\\
{\sl ${}^c$Instituto de F\'isica Te\'orica, (UAM/CSIC), Universidad Aut\'onoma de Madrid, Cantoblanco, E-28049 Madrid, Spain}\\
{\sl ${}^d$Instituto de F\'isica  Cantabria (CSIC-UC), E-39005 Santander, Spain}\\
{\sl ${}^e$Deutsches Elektronen-Synchrotron DESY, Notkestra{\ss}e 85, D-22607 Hamburg, Germany}\\
\def\thefootnote{\arabic{footnote}}
\setcounter{page}{0}
\setcounter{footnote}{0}
\end{center}
\vspace{2ex}
\begin{abstract}
{}
Various methods are used in the literature for predicting the lightest $\mathcal{CP}$-even Higgs boson mass in the Minimal Supersymmetric Standard Model (MSSM). Fixed-order diagrammatic calculations capture all effects at a given order and yield accurate results for scales of supersymmetric (SUSY) particles that are not separated too much from the weak scale. Effective field theory calculations allow a resummation of large logarithmic contributions up to all orders and therefore yield accurate results for a high SUSY scale. A hybrid approach, where both methods have been combined, is implemented in the computer code \FH. So far, however, at large scales sizeable differences have been observed between \FH\ and other pure EFT codes. In this work, the various approaches are analytically compared with each other
in a simple scenario in which all SUSY mass scales are chosen to be equal to each other.  Three main sources are identified that account for the major part of the observed differences. Firstly, it is shown that the scheme conversion of the input parameters that is commonly used for the comparison of fixed-order results is not
adequate for the comparison of results containing a series of higher-order
logarithms. Secondly, the treatment of higher-order terms arising from the
determination of the Higgs propagator pole is addressed. Thirdly, the effect
of different parametrizations in particular of the top Yukawa coupling in
the non-logarithmic terms is investigated. Taking into account all of these
effects, in the considered simple scenario very good agreement is found for
scales above 1~TeV between the results obtained using the EFT approach and
the hybrid approach of \FH. 

\end{abstract}

\newpage
\tableofcontents
\newpage
\def\thefootnote{\arabic{footnote}}

 
\section{Introduction}

The properties of the Higgs boson 
that has been discovered
by the ATLAS and CMS
collaborations at the CERN Large Hadron 
Collider~\cite{Aad:2012tfa,Chatrchyan:2012xdj} are compatible with those
predicted for the Higgs boson 
of the Standard Model (SM) at the present level of accuracy.
Despite this apparent
success of the SM, there are several open questions that cannot be
answered by the SM and ask for extended or alternative theoretical concepts.
Supersymmetry 
is one of best motivated frameworks for 
physics beyond the Standard Model (BSM), and in particular  
the Minimal Supersymmetric Standard Model (MSSM) is the 
most intensively studied scenario providing  precise predictions
for experimental 
phenomena
in the LHC era.

Apart from associating a superpartner to each SM degree of freedom, the MSSM
extends the Higgs sector of the SM by a second complex
doublet. Consequently, the MSSM
employs two Higgs-boson doublets, denoted by $H_1$ and $H_2$, with
hypercharges $-1$ and $+1$, respectively.   After minimizing the
scalar potential, the neutral components of $H_1$ and $H_2$ acquire
vacuum expectation values (vevs), $v_1$ and $v_2$. Without loss of
generality, one can assume that the vevs are real and non-negative, yielding
\begin{align}
v^2\equiv v_1^2+v_2^2, \quad
\tb\equiv v_2/v_1\,.
\end{align}
The two Higgs doublets in the MSSM accommodate five physical Higgs
bosons.  In lowest order these are the light and heavy $\cp$-even
Higgs bosons, $h$ and $H$, the $\cp$-odd Higgs boson, $A$, and two
charged Higgs bosons, $H^\pm$. Two parameters are required to describe 
the Higgs sector at the tree level (conventionally chosen
as $\tb$ and the mass $\MA$ of the $\cp$-odd Higgs particle);
masses and couplings, however,  are substantially affected by
higher-order contributions. 

Until now, experiments have not found direct evidence for 
supersymmetric (SUSY) 
particles. On the other hand, precision observables provide an indirect 
access to the MSSM parameter space from which significant
constraints on the allowed 
parameter regions can be obtained.
On top of the classical set of electroweak precision observables, the mass of the detected Higgs boson
constitutes an additional
important precision observable,
$M_h^{\rm{exp}}=125.09\pm0.24\gev$~\cite{Aad:2015zhl}.
If the measured value is associated with
the mass $M_h$ of the lightest $\cp$-even Higgs boson within the MSSM
(for a recent discussion of the viability of the interpretation in terms of
the heavy $\cp$-even Higgs boson $H$, see~\cite{Bechtle:2016kui}),
the comparison of the predicted value with the measurement constitutes an
important test of the model 
with high sensitivity to the SUSY mass
scales (see e.g.~\cite{Heinemeyer:2004ms,Heinemeyer:2004gx,Djouadi:2005gj}
for reviews).
In order to fully exploit the high precision of the experimental measurement 
for constraining the SUSY parameter space the accuracy of the theoretical
prediction for $M_h$ has to be improved very significantly.

So far, the full one-loop
corrections~\cite{Chankowski:1992er,Dabelstein:1994hb,Pierce:1996zz,Frank:2006yh},
dominant two-loop
corrections~\cite{Hempfling:1993qq,Casas:1994us,Carena:1995wu,Heinemeyer:1998kz,Heinemeyer:1998jw,Heinemeyer:1998np,Zhang:1998bm,Heinemeyer:1999be,Espinosa:1999zm,Espinosa:2000df,Degrassi:2001yf,Brignole:2001jy,Martin:2001vx,Brignole:2002bz,Martin:2002iu,Dedes:2003km,Martin:2002wn,Martin:2003it,Heinemeyer:2004xw,Martin:2004kr,Martin:2005eg,Heinemeyer:2007aq,Hollik:2014bua,Passehr:2017ufr}
and partial three-loop results \cite{Martin:2007pg,Harlander:2008ju,Kant:2010tf} for the light MSSM Higgs-boson mass have been calculated diagrammatically. Besides fixed-order
calculations, effective field theory (EFT) methods have been used to
resum large logarithmic contributions in case of a large mass hierarchy
between the electroweak and the SUSY scale~\cite{Giudice:2011cg,Draper:2013oza,Bagnaschi:2014rsa,Lee:2015uza,Vega:2015fna}. 
These EFT calculations, however, are less accurate for 
relatively low SUSY mass scales owing  
to terms suppressed by the SUSY scale(s) which 
correspond to higher-dimensional operators
in the EFT framework (see \cite{Bagnaschi:2017xid} for recent work in this direction).

In order to profit from the advantages of both methods -- high accuracy for
relatively low SUSY
scales in the case of the diagrammatic approach versus high accuracy for a high 
SUSY scale in the case of the EFT
approach -- a hybrid method combining both approaches has
been developed~\cite{Hahn:2013ria,Bahl:2016brp}, see 
also~\cite{Athron:2016fuq,Staub:2017jnp} for different implementations.
The method introduced in~\cite{Hahn:2013ria,Bahl:2016brp} has been implemented 
into the publicly available code
\FH~\cite{Heinemeyer:1998yj,Heinemeyer:1998np,Hahn:2009zz,Degrassi:2002fi,Frank:2006yh}
such that the fixed-order result is supplemented with higher-order
logarithmic contributions.

Comparisons between \FH and pure EFT codes in the
literature~\cite{Vega:2015fna,Athron:2016fuq,Staub:2017jnp} have revealed
non-negligible differences between the predicted values for $M_h$. 
In particular, deviations have been observed for large SUSY scales, 
where terms not captured in the EFT framework are supposed to be
negligible.
At first glance, such differences appear to be unexpected since the
resummation of logarithms included in \FH 
is at the same level of accuracy as in pure EFT calculations. 

In order to clarify the situation, it is
the purpose of this work to perform an in-depth comparison of the various approaches to explain 
the origin of the observed differences. For simplicity, we choose a single-scale scenario, 
\begin{align}\label{MSusyDef}
M_\text{soft} = \mu = \MA \equiv \msusy,
\end{align}
where $M_\text{soft}$ are the soft SUSY-breaking masses and $\mu$ is the Higgsino mass parameter. 
Furthermore, all parameters are assumed to be real, i.e.\ we work in the
$\cp$-conserving MSSM with real parameters.\footnote{Note that \FH works also with complex parameters including an interpolation of the resummation routines.} 
While the chosen single-scale scenario is particularly suitable for the EFT
approach, it should be noted that in realistic cases the actual task is to
provide the most accurate prediction (together with a reliable estimate of
the remaining theoretical uncertainties) for the Higgs-boson masses of the model
for a given SUSY mass spectrum which may contain a variety of SUSY scales.
We leave an investigation of such multi-scale scenarios for future work.

We shall explain that there are essentially three sources of the observed
differences. 
In a first step, we show that the usual scheme conversion of input
parameters is not suitable for 
the comparison of results containing a series of higher-order
logarithms. Such a scheme conversion can lead to large shifts corresponding to 
formally uncontrolled higher-order terms. 
Secondly, we analytically identify specific terms arising through the 
determination of the Higgs propagator pole which cancel with subloop
renormalization contributions in the irreducible self-energies of the
diagrammatic approach for a large SUSY scale. We develop an improved
treatment where unwanted effects from incomplete cancellations are avoided.
Thirdly, we show how different parametrizations of non-logarithmic terms can
explain remaining differences between the results of \FH and pure EFT codes
for high scales. 

The paper is organized as follows. In \Sec{HiggsPoleMassSection}, we review the different approaches with a
particular focus on how the Higgs pole mass is extracted. In
\Sec{CompSection}, we compare the results of the various approaches for
the Higgs pole mass to each other. In \Sec{FHwithDRinputSection}, we
discuss the issue of using $\DR$ input parameters as input of an OS
calculation. In \Sec{EFTcompSection}, we give a brief overview
about the levels of accuracy of the $\Mh$ evaluation
implemented in various codes. In \Sec{NumericalResultsSection}, we
present a numerical analysis showing the impact of the effects discussed
in the previous Sections and numerically compare \FH to other codes. 
The conclusions can be found in \Sec{ConclusionsSec}.
Two appendices provide additional details.


\section{Calculating the Higgs mass}\label{HiggsPoleMassSection}

In this Section, we shortly review how the pole mass of the lightest
$\mathcal{CP}$-even Higgs boson of the MSSM is calculated in a pure
diagrammatic calculation, in a pure EFT calculation, and in the hybrid
approach of \FH.


\subsection{Diagrammatic fixed-order calculation}\label{HiggsPoleMassDiagrammaticSection}

A well-established way to calculate corrections to the mass of the SM-like Higgs of the MSSM, as well as to the mass of the heavier 
\cp-even neutral Higgs boson and the charged Higgs boson,
is a fixed-order Feynman diagrammatic (FD)
calculation. The prediction is based on the calculation of
Higgs self-energies
involving contributions from SM particles, extra Higgs bosons, as
well as their corresponding superpartners. In this approach the
contributions from all sectors of the model and of all particles in the loop
can be incorporated at a given order. The mass effects of all particles in
the loop can be taken into account for any pattern of the mass spectrum. If
there is however a large splitting between the relevant scales, in
particular a large mass hierarchy between the electroweak and the scale of
some or all of the SUSY particles,
the fixed-order result will contain numerically large logarithms 
that can spoil the convergence of the perturbative expansion. 

In the MSSM with real parameters, after calculating the renormalized Higgs-boson self-energies, the
physical masses of the 
\cp-even Higgs bosons $h,H$ can be obtained
by finding the poles of their propagator matrix, whose inverse is given
by 
\begin{align}
\Delta_{hH}^{-1}=i 
\begin{pmatrix}
p^2 - m_h^2 + \hat\Sigma^\MSSM_{hh}(p^2) & \hat\Sigma^\MSSM_{hH}(p^2) \\
\hat\Sigma^\MSSM_{hH}(p^2) & p^2 - m_H^2 + \hat\Sigma^\MSSM_{HH}(p^2) 
\end{pmatrix},
\end{align}
where
$m_h$ ($m_H$) denotes the tree-level mass of the $h$ ($H$) boson and
$\hat\Sigma_{hh,hH,HH}$ are the corresponding self-energies. We
introduced the label ``MSSM'' to indicate that the corresponding self-energy contains SM-type contributions as well as non-SM contributions. 

Concerning the renormalization, we follow here the approach used in the program \FH. Accordingly, 
the circumflex $\hat{\mbox{ }}$  indicates
that the self-energies have been renormalized using the mixed 
on-shell (OS) and $\DR$-scheme of \cite{Frank:2006yh}. 
In particular, the $A$-boson mass is renormalized
on-shell, whereas the Higgs field renormalization and the
renormalization of $\tan\beta$ is performed using the $\DR$
scheme. 

The masses of the weak gauge bosons ($M_Z$, $M_W$) and the electromagnetic
charge $e$ are renormalized on-shell, and the tadpole renormalization is carried out such that the tadpole contributions are cancelled by their
respective counterterms. 
The OS vev is a dependent quantity, which is given in terms of the OS values of the observables $M_W$, $\SW$ and $e$ by
\begin{align}
v_\OS^2 = \frac{2 \SW^2 M_W^2}{e^2} ,
\label{eq:vOS}
\end{align}
where $\SW$ denotes the sine of the weak mixing angle. The renormalization of this quantity at the one-loop level
is therefore given in terms of the OS
counterterms of $M_W$, $\SW$ and~$e$,
\begin{align}\label{eq:vevCTnodeltaZ}
\frac{2 \SW^2 M_W^2}{e^2} \to 
&\frac{2 \SW^2 M_W^2}{e^2}\left\{ \frac{\delta\MW^2}{\MW^2}+\frac{\CW^2}{\SW^2}\left(\frac{\delta\MZ^2}{\MZ^2}-\frac{\delta \MW^2}{\MW^2}\right) -\frac{\delta e^2}{e^2} \right\},
\end{align}
where $\delta M_{W,Z}^2$ are the mass counterterms of the
$W$ and $Z$ bosons, respectively, and $\delta e^2$ is the counterterm
of the electromagnetic charge ($\CW^2=1-\SW^2$). 
Motivated by the fact that the renormalization of the vev receives
a contribution from the field renormalization of the Higgs doublet, we 
identify the counterterm given in \Eq{eq:vevCTnodeltaZ} with 
$\delta v_\OS^2/v_\OS^2 + \delta Z_{hh}$, where
$\delta Z_{hh}$ is the field renormalization counterterm of the SM-like
Higgs field fixed in the $\DR$ scheme.\footnote{Here, we already implictly assume the decoupling limit ($M_A\gg M_Z$) in the sense that we identify the $h$ boson as the SM-like Higgs.} Accordingly, 
the OS counterterm  of the vev defined in this way reads
\begin{align}\label{eq:vevCT}
\frac{\delta v_\OS^2}{v_\OS^2} = \frac{\delta
\MW^2}{\MW^2}+\frac{\CW^2}{\SW^2}\left(\frac{\delta
\MZ^2}{\MZ^2}-\frac{\delta \MW^2}{\MW^2}\right)-\frac{\delta e^2}{e^2} 
- \delta Z_{hh}.
\end{align}
The results for the self-energies in \FH\ have been reparametrized in terms 
of the Fermi constant $G_F$ instead of the electric charge $e$. 
The corresponding vev $v_{G_F}$ is related to $v_\OS$ via
\begin{align}\label{vevDef}
v_\OS^2  = v_{G_F}^2(1+\Delta r) \text{ with }v_{G_F}^2= \frac{1}{2\sqrt{2}G_F} .
\end{align}
MSSM predictions for the quantity 
$\Delta r$ 
can be found in 
\cite{Chankowski:1993eu,Heinemeyer:2006px,Heinemeyer:2013dia,Stal:2015zca}.
The effect of this reparametrization in the one-loop self-energies is formally of two-loop order.

Furthermore (in the default choice), the stop sector is renormalized using the OS
scheme, which is defined by applying on-shell conditions for the
respective masses: the top-quark mass $M_t$, and the top-squark masses
$\Mstope$ and $\Mstopz$. A fourth renormalization condition fixes the
mixing of the stops and can be identified with a condition for the top-squark
mixing angle.

Employing this scheme, in \FH\ the full one-loop
corrections to the Higgs self-energies as well as two-loop corrections
of \order{\alt\als,\alb\als,\alt^2,\alt\alb,\alb^2} are implemented~\cite{Heinemeyer:1998yj,Heinemeyer:1998np,Degrassi:2001yf,Brignole:2001jy,Brignole:2002bz,Degrassi:2002fi,Dedes:2003km,Heinemeyer:2004xw,Frank:2006yh,Heinemeyer:2007aq,Hahn:2009zz,Hollik:2014bua}. 
While those two-loop corrections in the gauge-less limit have been obtained
for vanishing external momentum, there is futhermore an option to
incorporate the momentum dependence of the corrections at
\order{\alt\als}~\cite{Borowka:2014wla,Borowka:2015ura} (see also \cite{Degrassi:2014pfa}).
Finding the (complex) poles for the case where $\cp$ conservation is assumed corresponds to solving the equation 
\begin{align}\label{FullHiggsPoleEq}
{}&\left(p^2 - m_h^2 + \hat\Sigma_{hh}^\MSSM(p^2)\right)\left(p^2 - m_H^2 + \hat\Sigma^\MSSM_{HH}(p^2)\right)-\left(\hat\Sigma^\MSSM_{hH}(p^2)\right)^2 = 0.
\end{align}
In the decoupling limit, $\MA\gg M_Z$, the physical mass of the lightest
Higgs boson can approximately be obtained as solution of the simpler equation
\begin{align}\label{HiggsPoleEq}
p^2 - m_h^2 + \hat\Sigma^\MSSM_{hh}(p^2) = 0
\end{align}
up to corrections from the $hH$ and $HH$ self-energies, which are suppressed
by powers of $M_A$.  In the following discussion we will for simplicity use \Eq{HiggsPoleEq}
for determining the pole of the propagator and we will furthermore neglect
the imaginary parts of the self-energies. In \FH\ the complex poles of the
propagator are obtained from the full propagator matrix, taking into account the real and imaginary parts of the Higgs-boson self-energies.

Solving \Eq{HiggsPoleEq} iteratively for the case where imaginary parts are
neglected yields an expression for the Higgs pole mass,
\begin{align}\label{FDpoleMass}
(M_h^2)_\text{FD} ={}& m_h^2 - \hat\Sigma_{hh}^\MSSM(m_h^2)
+\hat\Sigma_{hh}^{\MSSM\prime}(m_h^2) \hat\Sigma_{hh}^\MSSM(m_h^2) + \ldots \, ,
\end{align}
where the prime denotes the derivative of the self-energy with respect
to the momentum squared. The ellipsis stands for terms involving
higher-order 
derivatives and products of differentiated self-energies. 
In \App{p2TermsApp} we provide a formula from which these terms can be derived recursively.
The Higgs pole mass at a given order is obtained from \Eq{FDpoleMass} via a loop expansion to the appropriate order.


\subsection{Effective Field Theory calculation}\label{HiggsPoleMassEFTSection}

Another approach to calculate the mass of the SM-like Higgs boson in the MSSM is
using effective field theory (EFT) methods. These allow the resummation of large logarithmic contributions, so that higher-order contributions beyond 
the order of fixed-order diagrammatic calculations can be incorporated. 
Without including higher dimensional operators in
the effective Lagrangian, contributions suppressed by a heavy scale are
however not captured.

In the simplest EFT framework, all SUSY particles are integrated out
from the full theory at a common mass scale $\msusy$. Below $\msusy$
the SM remains as the low-energy  EFT. 
The couplings of the EFT are determined by
matching to the MSSM at the scale $\msusy$. 
In the case of the SM as the EFT\footnote{In case of $M_A\sim M_t$ 
the effective theory is a Two-Higgs-Doublet model and not the SM,
see \cite{Lee:2015uza}.}
below $\msusy$ 
this concerns only the effective Higgs self-coupling $\lambda$,
all the other couplings are fixed by matching them to observables
at the low-energy scale.
Renormalization group equations (RGEs) are used to
correlate the couplings at the high scale $\msusy$ and the
low scale, typically chosen to be the OS top
mass $M_t$ (or $M_Z$).  

The effective Higgs self coupling $\lambda(M_t)$ obtained 
from the matched $\lambda(\msusy)$ determines the $\MS$ mass of 
the SM Higgs boson at the scale $M_t$ via
\begin{align}
\label{eq:runningmass} 
(m_h^{\MS,\text{SM}})^2 = 2\, \lambda(M_t)\, v_\MS^2 \, ,
\end{align}
with the $\MS$ vev  (at the scale $M_t$). The $\MS$ vev can be related to 
the on-shell vev via the finite part 
of $\delta v_\OS^2$ defined in~\Eq{eq:vevCT},
\begin{align}
\label{eq:vevrel}
  v_\MS^2 = v_\OS^2 + \delta v_\OS^2 \Big|_\text{fin} \, .
\end{align}
It should be noted that since the quantity in \Eq{eq:runningmass} is the SM $\MS$ vev, in \Eq{eq:vevrel} only SM-type contributions have to be 
considered in $\delta v_\OS^2$.

Getting from the running mass (\ref{eq:runningmass}) to
the physical Higgs mass one has to solve the 
pole equation for the Higgs-boson propagator,
\begin{align}
p^2 - (m_h^{\MS,\text{SM}})^2 + \tilde\Sigma_{hh}^\SM(p^2) = 0,
\end{align}
involving the renormalized SM Higgs boson self-energy
(denoted by a tilde)  
\begin{align}
\tilde\Sigma_{hh}^\SM(p^2) = 
   \Sigma_{hh}^{\SM}(p^2)\Big|_\text{fin}
  - \frac{1}{\sqrt{2} v_\MS}\, T_h^\SM\Big|_\text{fin} \, ,
\end{align}
which is renormalized accordingly  
in the $\MS$ scheme at the scale $M_t$ 
but with the Higgs tadpoles renormalized to zero,
i.e.\ the tadpole counterterm is chosen to cancel 
the sum of the  tadpole diagrams, $T_h^\SM$, 
for the Higgs~field,  
\begin{align}
\delta T_h^\SM = - T_h^\SM \, .
\end{align}
With all these ingredients, the
Higgs pole mass is now obtained as the solution of the equation
\begin{align}
M_h^2 = 2 \lambda(M_t) v_\MS^2\,  -\, \tilde\Sigma_{hh}^\SM(M_h^2).
\end{align}
Expanding the Higgs self-energy perturbatively around  the tree-level mass $m_h^2$  
of the MSSM yields 
\begin{align}\label{EFTpoleMass}
(M_h^2)_\EFT ={}& 2v_\MS^2\lambda(M_t)  - \tilde\Sigma_{hh}^\SM(m_h^2) -\,\tilde \Sigma_{hh}^{\SM\prime}(m_h^2)\cdot\left[2
  v_\MS^2\lambda(M_t) - \tilde\Sigma_{hh}^\SM(m_h^2) - m_h^2\right]+\, \cdots \, ,  
\end{align}
where the ellipsis indicates higher-order terms in the expansion.

We discuss the current status of EFT calculations in \Sec{EFTcompSection}.


\subsection{Hybrid calculation}\label{HiggsPoleMassHybridSection}

In \FH, the fixed-order approach is combined with the EFT
approach in order to supplement the full diagrammatic result with leading higher-order contributions~\cite{Hahn:2013ria,Bahl:2016brp}. 
The logarithmic contributions resummed using the EFT approach are 
incorporated into
\Eq{HiggsPoleEq},
\begin{align}\label{HiggsPoleEqDelta}
p^2-m_h^2+\hat\Sigma^\MSSM_{hh}(p^2) +\Delta\hat\Sigma_{hh}^2 = 0.
\end{align}
The quantity $\Delta\hat\Sigma_{hh}$ contains all logarithmic contributions obtained via
the EFT approach as well as subtraction terms compensating the logarithmic
terms already present in the diagrammatic fixed-order result for $\hat\Sigma^\MSSM_{hh}$,
\begin{align}\label{subtractionterms_Eq}
\Delta\hat\Sigma_{hh}^2 = -\big[2 v_\MS^2\lambda(M_t)\big]_\text{log} - \big[\hat\Sigma_{hh}^\MSSM(m_h^2)\big]_\text{log}.
\end{align}
The subscript `log' indicates that we take only logarithmic
contributions into account. Note that in $\hat\Sigma_{hh}^\MSSM(m_h^2)$ the logarithms appear only explicitly when expanding in $v/\msusy$. For more details on the combination of the fixed-order and the EFT result, we refer to \cite{Bahl:2016brp,Hahn:2013ria}. 

Plugging the expression for $\Delta\hat\Sigma_{hh}$ into \Eq{HiggsPoleEqDelta}, we obtain for the physical Higgs mass
\begin{align}
(M_h^2)_\text{FH} ={}& m_h^2  - \hat\Sigma^\MSSM_{hh}(M_h^2) + \big[2v_\MS^2\lambda(M_t)\big]_\text{log} + \big[\hat\Sigma^\MSSM_{hh}(m_h^2)\big]_\text{log} = \nonumber\\
={}&m_h^2 + \big[2v_\MS^2\lambda(M_t)\big]_\text{log} - \big[\hat\Sigma^\MSSM_{hh}(m_h^2)\big]_\text{nolog}  \nonumber\\
&-
\hat\Sigma^{\MSSM\prime}_{hh}(m_h^2)\left(\big[2v_\MS^2\lambda(M_t)\big]_\text{log}- \big[\hat\Sigma^\MSSM_{hh}(m_h^2)\big]_\text{nolog}\right) \nonumber\\
{}&+ \ldots \; .
\label{FHpoleMass}
\end{align}
We use the label `nolog' to indicate that we take only terms not involving
large logarithms into account for the labelled quantity. We again would like
to stress that
the large logarithms (and thereby the meant non-logarithmic terms) appear
only explicitly in $\hat\Sigma_{hh}^\MSSM(m_h^2)$ when expanding in
$v/\msusy$.

Before comparing the various approaches in depth, we also shortly comment on the renormalization scheme conversion needed for the combination of the fixed-order and the EFT calculation. As mentioned before, in \FH (in the default choice) the stop sector is renormalized using the OS
scheme. In contrast, in the EFT calculation, i.e.\ the 
calculation of $\lambda(M_t)$, all SUSY parameters enter in
$\DR$-renormalized form. As argued in \cite{Bahl:2016brp}, it is
sufficient to convert only the stop mixing parameter $\Xt$ using only
the one-loop large logarithmic terms, 
\begin{align}\label{XtEFT}
\Xt^{\DR,\EFT} ={}& \Xt^\OS\left[1 + \left(\frac{\als}{\pi} -\frac{3 \alt}{16\pi}(1-\Xt^2/M_S^2)\right)\ln \frac{M_S^2}{M_t^2}\right],
\end{align}
where $M_S^2 = \Mstope\Mstopz$, $\als = g_3^2/(4\pi)$ (with $g_3$ being the strong gauge
coupling) and $\alt = y_t^2/(4\pi)$ (with $y_t$ being the top Yukawa
coupling).


\section{Comparison of the different approaches}\label{CompSection}

In the following we will discuss the differences between the
various approaches. It is obvious from the discussion of the previous
section that the diagrammatic fixed-order result and the pure EFT result
differ by higher-order logarithmic terms that are contained in the EFT
result but not in the diagrammatic fixed-order result as well as by
non-logarihmic terms that are contained in the 
diagrammatic fixed-order result but not in the pure EFT result. In the
hybrid approach the diagrammatic fixed-order result is supplemented by the
higher-order logarithmic terms obtained by the EFT approach. We focus in the
following on the comparison between the hybrid approach and the pure EFT
result. In the present section we leave aside issues related
to the used renormalization schemes, which will be addressed in
\Sec{FHwithDRinputSection}. 

While the hybrid approach and the pure EFT approach both incorporate the
higher-order logarithmic terms obtained by the EFT approach, this does not
necessarily imply that all logarithmic terms in the two results are the
same. This is due to the fact that the determination of the Higgs-boson mass from
the pole of the progagator within the hybrid approach 
is performed in the full model (in the example considered here the MSSM,
incorporating loop contributions from all SUSY particles), while in the EFT
approach it is determined in the effective low-scale model (in the
considered example the SM). We will demonstrate below that the determination
of the propagator pole in the hybrid approach generates logarithmic terms 
beyond the ones contained in the EFT approach at
the two-loop level and beyond which actually cancel in the limit of a heavy
SUSY scale with contributions from the subloop renormalization. This
cancellation is explicitly demonstrated at the two-loop level.
We will furthermore discuss the difference in non-logarithmic terms between
the results of the hybrid and the EFT approach.


\subsection{Higher-order logarithmic terms from the determination of the pole
of the propagator}\label{LogCompSection}

In the EFT approach where the Higgs boson mass is determined as the pole of the
propagator in the SM as the effective low-scale model, while the SUSY
particles have been integrated out, 
the logarithmic terms are given by (see \Eq{EFTpoleMass})
\begin{align}
(M_h^2)_\EFT^\text{log} ={}& \left[2v_\MS^2\lambda(M_t)\right]_\text{log} -
\tilde\Sigma_{hh}^{\SM\prime}(m_h^2)
\left[2v_\MS^2\lambda(M_t)\right]_\text{log} + \ldots .
\end{align}
The logarithmic terms contained in the result of
the hybrid approach implemented in \FH are given by
(see \Eq{FHpoleMass}) 
\begin{align}
(M_h^2)_\text{FH}^\text{log} =&{} \left[2v_\MS^2\lambda(M_t)\right]_\text{log}+ \left[\hat\Sigma^{\MSSM\prime}_{hh}(m_h^2)\right]_\text{log}\left[\hat\Sigma_{hh}^\MSSM(m_h^2)\right]_\text{nolog} \nonumber\\
&- \hat\Sigma^{\MSSM\prime}_{hh}(m_h^2)
\left[2v_\MS^2\lambda(M_t)\right]_\text{log}  + \ldots .
\end{align}
In the decoupling limit ($\msusy = M_A \gg M_t$, where in particular the
  light $\cp$-even Higgs boson has SM-like couplings), we can split up the
MSSM Higgs self-energy into a SM part and a non-SM part, 
\begin{align}
\hat\Sigma^{\MSSM}_{hh}(m_h^2) = \hat\Sigma_{hh}^\SM(m_h^2) + \hat\Sigma_{hh}^\nonSM(m_h^2).
\end{align}
In the mixed OS/$\DR$ scheme of the full diagrammatic calculation, the Higgs
field renormalization constants are fixed in the $\DR$ scheme. For scalar propagators, there is no difference between the $\DR$ and the $\MS$ scheme at the one-loop level. Consequently,
\begin{align}
\hat\Sigma_{hh}^{\SM\prime}(m_h^2) = \tilde\Sigma_{hh}^{\SM\prime}(m_h^2) 
\end{align}
holds.

Using this relation, we obtain for the difference between the higher-order 
logarithmic terms from the determination of the pole of the propagator
obtained in the EFT and the hybrid approach 
\begin{align}
\Delta^\text{log} \equiv{}& (M_h^2)_\text{FH}^\text{log}-(M_h^2)_\EFT^\text{log} 
                                                                  \nonumber\\
={}& \left[\hat\Sigma^{\nonSM\prime}_{hh}(m_h^2)\right]_\text{log}
   \left[\hat\Sigma_{hh}^\MSSM(m_h^2)\right]_\text{nolog}   - \hat\Sigma^{\nonSM\prime}_{hh}(m_h^2)
   \left[2v_\MS^2\lambda(M_t)\right]_\text{log} + \ldots  
\label{p2TermsLog} \\
=&: \Delta_{p^2}^\text{log} . \nonumber 
\end{align}
Since this difference, which is of two-loop order and beyond,
results only from the momentum dependence of the non-SM contributions to the
Higgs self-energy, we call it $\Delta_{p^2}^\text{log}$ in the following. We give analytic expressions for $\Delta_{p^2}^\text{log}$ in \App{p2TermsApp}.

In \Sec{p2TermAtTwoLoopSection} we will demonstrate at the two-loop level
that in the limit of a heavy SUSY scale the quantity
$\Delta_{p^2}^\text{log}$ consisting of ``momentum-dependent non-SM
contributions'' as given in \Eq{p2TermsLog} cancels out with contributions
of the Higgs self-energy's subloop renormalization. Before we address this issue we first
compare the non-logarithmic terms in the two approaches.

\subsection{Non-logarithmic terms}\label{NonLogCompSection}


In the EFT approach, the non-logarithmic terms are given by (see \Eq{EFTpoleMass})
\begin{align}
(M_h^2)_\EFT^\text{nolog} =&{} \left[2v_\MS^2\lambda(M_t)\right]_\text{nolog} - \tilde\Sigma_{hh}^\SM(m_h^2) -
\tilde\Sigma_{hh}^{\SM\prime}(m_h^2)\Big(\left[2v_\MS^2\lambda(M_t)\right]_\text{nolog} - \tilde\Sigma_{hh}^\SM(m_h^2)- m_h^2\Big)+
\ldots \; . \label{eq:EFTnonlog}
\end{align} 

By construction, 
all non-logarithmic terms contained in the result of the hybrid approach
originate from the fixed-order diagrammatic calculation (see 
\Eq{FHpoleMass}),
\begin{align}\label{MhFHnolog}
(M_h^2)_\text{FH}^\text{nolog} ={}& m_h^2 -
\left[\hat\Sigma_{hh}^\MSSM(m_h^2)\right]_\text{nolog} +\left[\hat\Sigma_{hh}^{\MSSM\prime}(m_h^2)\right]_\text{nolog}\left[\hat\Sigma_{hh}^\MSSM(m_h^2)\right]_\text{nolog}+\ldots \; .
\end{align}
In this way one- and two-loop terms that are suppressed by the SUSY scale, 
$\Delta^\text{nolog}_{v/\msusy}$, are included in the result of the hybrid
approach. Terms of this kind
would result from higher-dimensional
operators in the EFT approach. Those terms that are included in the 
hybrid result as implemented in \FH\ but not in the publicly available pure EFT
results constitute an important source of difference between the
corresponding results, which is expected to be sizeable if some or all SUSY
particles are relatively light (see also \cite{Bagnaschi:2017xid} for a
recent discussion of contributions of this kind in the EFT approach). 
It should be noted that in general terms of 
\order{v/\msusy} also originate 
from solving the full pole mass equation, \Eq{FullHiggsPoleEq}, rather than
the approximated one, \Eq{HiggsPoleEq}.  

At zeroth order in $v/\msusy$,
the non-logarithmic terms of the EFT approach contained in $\lambda(M_t)$ in
\Eq{eq:EFTnonlog} agree with the non-SM contributions in \Eq{MhFHnolog}. 
They result from the threshold
corrections at the matching scale $\msusy$.
These threshold corrections are
so far only known fully at the one-loop order. At the two-loop order
only the \order{\als\alt,\alt^2} corrections are implemented in publicly available codes so far.\footnote{Two-loop corrections controlled by the bottom and tau Yukawa couplings have recently been derived in \cite{Bagnaschi:2017xid}.} 
Thus, those terms in
$\left[\hat\Sigma_{hh}^{\nonSM\prime}(m_h^2)\right]_\text{nolog}\left[\hat\Sigma_{hh}^\MSSM(m_h^2)\right]_\text{nolog}$
being not of \order{\alt^2} are not present in $(M_h^2)_\EFT$. At
higher orders, all terms involving a derivative of
$\hat\Sigma_{hh}^\nonSM$ are affected. As we will demonstrate in the following section, also the non-logarithmic
non-SM contributions arising from the determination of the pole of the
propagator cancel out with contributions
of the subloop renormalization in the limit of a high SUSY scale.

Apart from these terms and from the non-logarithmic terms of 
\order{v/\msusy} discussed above,
$\Delta^\text{nolog}_{v/\msusy}$, 
a further difference between the hybrid approach and the EFT aproach
is due to the parametrization of the non-logarithmic terms. 
In the EFT approach all low-scale parameters are $\MS$ quantities. The
results of \FH, on the other hand, are expressed in terms of physical, i.e.\
on-shell, parameters. For the top-quark mass both the results expressed in
terms of the pole mass, $M_t$, and the running mass at the scale $M_t$, 
$\overline{m}_t(M_t)$ 
(see \cite{Williams:2011bu} for details on the involved reparametrization)
have been implemented (the applied renormalization
schemes for SUSY parameters will be discussed below).
The Higgs vev is a dependent
quantity in \FH\ which is expressed in terms of the physical observables
$\MW$, $\SW$ and $e$ according to \Eq{eq:vOS} (where $e$ is furthermore
reparametrized in terms of the Fermi constant, see \Eq{vevDef}). 
Accordingly, if choosing low-energy SM parameters to express the EFT result, the non-logarithmic terms in this result are parametrized
in terms of the $\MS$ quantities 
$\overline{m}_t(M_t)$ and $v_\MS(M_t)$, while depending on the option chosen
for the top-quark mass the non-logarithmic terms in \FH\ are expressed in terms of 
either $\overline{m}_t(M_t)$ and $v_{G_F}$ or $M_t$ and $v_{G_F}$. Those
parametrizations differ from each other by higher-order terms. The observed
differences are therefore related to the remaining uncertainties of unknown
higher-order corrections.

It should be noted that also within the EFT approach there is a certain
freedom for choosing different parametrizations.
For instance, the threshold corrections at the matching scale can be
expressed in terms of the SM $\MS$ top Yukawa coupling or in terms
of the MSSM $\DR$ top Yukawa coupling. 

As a result, the deviations $\Delta^\text{nolog}$ between the
non-loga\-rithmic terms in the hybrid approach and the EFT approach arise from
the following sources,
\begin{align}\label{p2TermsNonLog}
\Delta^\text{nolog} \equiv{} & (M_h^2)_\text{FH}^\text{nolog}-(M_h^2)_\EFT^\text{nolog} = \nonumber\\
={} & \Delta^\text{nolog}_{v/\msusy} + \Delta^\text{nolog}_\text{para} +
\Delta^\text{nolog}_{p^2} .
\end{align}
Here $\Delta^\text{nolog}_{v/\msusy}$ are terms present in the hybrid
approach that would correspond to higher-dimensional operators in the EFT 
approach. The term $\Delta^\text{nolog}_\text{para}$
indicates the differences in the parametrization of the non-logarithmic
terms, and
\begin{align}
\Delta^\text{nolog}_{p^2}:={} &\left[\hat\Sigma^{\nonSM\prime}_{hh}(m_h^2)\right]_\text{nolog}\left[\hat\Sigma_{hh}^\MSSM(m_h^2)\right]_\text{nolog}\nonumber\\
&-\left[\hat\Sigma^{\nonSM\prime}_{hh}(m_h^2)\right]_\text{nolog}^{\order{\alt}}\left[\hat\Sigma_{hh}^\MSSM(m_h^2)\right]_\text{nolog}^{\order{\alt}} \nonumber\\
&+ \Big[\text{higher order terms involving}(\partial/\partial p^2)^n \hat\Sigma_{hh}^\nonSM, n\geq 1\Big] \label{p2NonLogDef}
\end{align}
are terms arising from the different determination of the propagator poles,
as discussed above.

\subsection{Terms arising from the determination of the propagator pole 
at the two-loop level}
\label{p2TermAtTwoLoopSection}

We saw in \Sec{LogCompSection} and \Sec{NonLogCompSection} that the
different
determination of the propagator pole in the hybrid approach and the EFT 
approach gives rise to both logarithmic and non-logarithmic contributions in
which the expressions given for the two approaches in the previous sections
differ from each other. We will now explicitly demonstrate at the two-loop
level that those differences in fact cancel out in the limit of a heavy SUSY
scale if all the relevant terms at this order are taken into account. 

As a first step, we write down
the correction to $M_h^2$, derived by an explicit diagrammatic
calculation. At strict two-loop order, we obtain 
\begin{align}\label{TwoLoopFDresult}
(M_h^2)_\text{FD} ={} & m_h^2 - \hat\Sigma_{hh}^{\MSSM,(1)}(m_h^2) - \hat\Sigma_{hh}^{\MSSM,(2)}(m_h^2) \nonumber\\
&+ \left(\hat\Sigma_{hh}^{\nonSM,(1)\prime}(m_h^2)+ \hat\Sigma_{hh}^{\SM,(1)\prime}(m_h^2)\right)\hat\Sigma_{hh}^{\MSSM,(1)}(m_h^2).
\end{align}
The superscripts indicate the loop-order of the corresponding
self-energy.%
\footnote{
In our discussion here we treat the two-loop self-energy as the full result
containg all contributions that appear at this order. The specific
approximations that have been made at the two-loop level in \FH\ will be
discussed below.
}  

We obtain the renormalized two-loop self-energy from the unrenormalized one via
\begin{align}
\hat\Sigma_{hh}^{\MSSM,(2)}(m_h^2) ={}& \Sigma_{hh}^{\MSSM,(2)}(m_h^2) + (\text{two-loop counterterms}) + (\text{subloop-ren.}).
\end{align}
The subloop-renormalization can be derived from the one-loop self-energy
via a counterterm-expansion. Expressing all couplings appearing in the one-loop
self-energy through masses divided by $v_{G_F}$ (for the remainder of this
section we drop the subscript ``$G_F$'', i.e.\ we use the shorthand 
$v \equiv v_{G_F}$), we can write
\begin{align}\label{TLsubren}
(\text{su}&\text{bloop-ren.}) =\nonumber\\
={}& (\delta v^2)^\MSSM \frac{\partial}{\partial v^2}\hat\Sigma_{hh}^{\MSSM,(1)}(m_h^2) + \sum_i (\delta m_i)^\MSSM \frac{\partial}{\partial m_i}\hat\Sigma_{hh}^{\MSSM,(1)}(m_h^2) +\text{(field ren.)}. =\nonumber\\
={}& -\frac{(\delta v^2)^\MSSM}{v^2} \hat\Sigma_{hh}^{\MSSM,(1)}(m_h^2)+ \sum_i (\delta m_i)^\MSSM \frac{\partial}{\partial m_i}\hat\Sigma_{hh}^{\MSSM,(1)}(m_h^2) +\text{(field ren.)}, 
\end{align}
where we used in the last line that
$\hat\Sigma_{hh}^{\MSSM ,(1)}\propto 1/v^2$ if all couplings are expressed by the respective mass divided by $v$. 

We are interested in terms involving the finite parts of the derivative
of the Higgs self-energy, i.e.\ terms which could potentially
cancel the term proportional to
$\hat\Sigma_{hh}^{\nonSM,(1)\prime}(m_h^2)$ in
\Eq{TwoLoopFDresult}. At first sight it would seem that terms of this kind
could arise from an on-shell field renormalization of the Higgs field. It is
well-known, however, that those field renormalization constants drop out of
the prediction of the mass parameter order by order in perturbation theory
(in \FH, a $\DR$ renormalization is employed for the Higgs fields).
Also the mass counterterms as well as the genuine two-loop counterterms do not contribute terms that are 
proportional to $\hat\Sigma_{hh}^{\nonSM,(1)\prime}(m_h^2)$.
The only remaining term is the vev counterterm. 
According to \Eq{eq:vevCT} and \Eq{vevDef} it is given at the one-loop level
by, having the same form in the SM
and the MSSM, 
\begin{align}\label{OSvev}
\frac{\delta v^2}{v^2} ={}& \frac{\delta
\MW^2}{\MW^2}+\frac{\CW^2}{\SW^2}\left(\frac{\delta
\MZ^2}{\MZ^2}-\frac{\delta \MW^2}{\MW^2}\right)-\frac{\delta e^2}{e^2} +\Delta r - \delta Z_{hh}.
\end{align}
The renormalization constant
$\delta Z_{hh}$ represents within the MSSM the $\DR$ field
renormalization constant of the SM-like Higgs field, while in the SM it is 
understood to be the $\MS$ field renormalization constant of the Higgs
field.

We verified by explicit calculation that in the limit of a large SUSY scale
the following relation holds
\begin{align}\label{MSSMvevSMvev}
\frac{(\delta v^2)^\MSSM}{v^2} ={}&  \frac{(\delta v^2)^\SM}{v^2} - \hat\Sigma_{hh}^{\nonSM,(1)\prime}(m_h^2) + \order{v/\msusy}.
\end{align}
Using this relation, we can rewrite the two-loop self-energy (omitting terms of \order{v/\msusy}),
\begin{align}
\hat\Sigma_{hh}^{\MSSM,(2)}(m_h^2) ={}& \hat\Sigma_{hh}^{\MSSM,(2)}(m_h^2)\Big|_{(\delta v^2)^\MSSM\to(\delta v^2)^\SM}+ \hat\Sigma_{hh}^{\nonSM,(1)\prime}(m_h^2)\hat\Sigma_{hh}^{\MSSM,(1)}(m_h^2),
\end{align}
where the subscript `$(\delta v^2)^\MSSM\to(\delta v^2)^\SM$' is used to indicate that the MSSM vev counterterm, appearing in the subloop renormalization, is replaced by its SM counterpart.

Plugging this expression back into \Eq{TLsubren} and \Eq{TwoLoopFDresult}, we obtain
\begin{align}
(M_h^2)_\text{FD} ={} & m_h^2 - \hat\Sigma_{hh}^{\MSSM,(1)}(m_h^2) \nonumber\\
&- \left(\hat\Sigma_{hh}^{\MSSM,(2)}(m_h^2)\Big|_{(\delta
v^2)^\MSSM\to(\delta v^2)^\SM}+\hat\Sigma_{hh}^{\nonSM,(1)\prime}(m_h^2)\hat\Sigma_{hh}^{\MSSM,(1)}(m_h^2)\right)\nonumber\\
&+ \left(\hat\Sigma_{hh}^{\nonSM,(1)\prime}(m_h^2)+ \hat\Sigma_{hh}^{\SM,(1)\prime}(m_h^2)\right)\hat\Sigma_{hh}^{\MSSM,(1)}(m_h^2)=\nonumber\\
={} & m_h^2 - \hat\Sigma_{hh}^{\MSSM,(1)}(m_h^2)- \hat\Sigma_{hh}^{\MSSM,(2)}(m_h^2)\Big|_{(\delta v^2)^\MSSM\to(\delta v^2)^\SM}+ \hat\Sigma_{hh}^{\SM,(1)\prime}(m_h^2)\hat\Sigma_{hh}^{\MSSM,(1)}(m_h^2).
\end{align}
We observe that the corresponding subloop renormalization term cancels
in \Eq{TwoLoopFDresult} the term 
$\hat\Sigma_{hh}^{\nonSM,(1)\prime}(m_h^2)$ involving the 
non-SM contributions to the Higgs self-energy by which the determination of
the propagator pole in the hybrid approach differs from the EFT approach.

The origin of \Eq{MSSMvevSMvev} is the different normalization of the
SM-like MSSM Higgs doublet~$\Phi_\MSSM$ and the SM Higgs doublet~$\Phi_\SM$.
Comparing the derivative of the two-point function, appearing in the LSZ
factor of amplitudes with external Higgs fields,%
\footnote{It should be noted that such an LSZ factor enters in the EFT
approach via the matching condition at the high scale.}
we obtain in the limit of a heavy SUSY scale
\begin{align}
&\Phi_\MSSM\left(1 +
\frac{1}{2}\hat\Sigma_{hh}^{\MSSM,(1)\prime}(m_h^2)\right) =\Phi_\SM\left(1
+ \frac{1}{2}\hat\Sigma_{hh}^{\SM,(1)\prime}(m_h^2)\right) ,
\end{align}
or equivalently
\begin{align}
\Phi_\MSSM = \Phi_\SM\left(1 - \frac{1}{2}\hat\Sigma_{hh}^{\nonSM,(1)\prime}(m_h^2)\right). 
\end{align}
Expressed in terms of a relation between the counterterms of the vevs, this
implies \Eq{MSSMvevSMvev}.

While as mentioned above the Higgs field renormalization constant drops out
in the Higgs mass prediction order by order, it is nevertheless noteworthy
that the introduction of an OS field renormalization constant would lead to
\begin{align}
\hat\Sigma^{\MSSM\prime}_{hh}(m_h^2)\big|_{\delta Z_h^\OS} = 0
\end{align}
and
\begin{align}
(\delta v^2)^\MSSM\big|_{\delta Z_h^\OS} &=  (\delta v^2)^\SM\big|_{\delta Z_h^\OS} ,
\end{align}
implying that no terms involving
$\hat\Sigma_{hh}^{\nonSM\prime}$ appear in the subloop renormalization at
the two-loop level. 

While we have demonstrated this cancellation at the two-loop level, it is to
be expected that it would also occur at higher orders. Explicit formulas for
higher-order terms of this kind are given in \App{p2TermsApp}. While the described cancellation occurs at the full two-loop level, only
partial cancellations occur between the full one-loop self-energy times its
derivative and the two-loop self-energy if for the latter certain
approximations are made.

In \FH, the two-loop self-energies are derived in the gaugeless limit (i.e.,
two-loop corrections of $\mathcal{O}(\alt\als,$ $\alb\als,\alt^2,\alt\alb,\alb^2)$ are incorporated
\cite{Degrassi:2001yf,Brignole:2001jy,Brignole:2002bz,Dedes:2003km,Heinemeyer:2007aq,Hollik:2014bua}),\footnote{The recent results of \cite{Passehr:2017ufr} for the \order{\alt\alb,\alb^2} corrections in the
general case of complex parameters will be implemented into \FH.} and by default the external momentum of the two-loop graphs is neglected. There is, however, an option to include momentum dependence at
\order{\alt\als} (see \cite{Borowka:2014wla,Borowka:2015ura}). Accordingly, all \order{\alt^2,\alt\alb,\alb^2}
non-SM terms arising through the determination of the propagator pole at the two-loop level are cancelled in
the limit of a large SUSY scale by
corresponding subloop renormalization contributions
within the diagrammatic calculation
(the determination of the propagator pole obviously does not give rise to
terms of \order{\alt\als,\alb\als}).
In previous versions of \FH, we have
already taken care when constructing the subtraction terms according to 
\Eq{subtractionterms_Eq} that we do not subtract logarithmic contributions 
that are needed for the cancellation with the corresponding terms arising
from the determination of the propagator poles. For terms arising through
the determination of the propagator pole beyond 
\order{\alt^2,\alt\alb,\alb^2}, however, so far the cancellation
in the limit of a large SUSY scale did not occur because the corresponding
contributions in the irreducible self-energies 
at the two-loop level and beyond are not incorporated. 
In order to avoid unwanted effects from an incomplete cancellation, we have
removed the uncompensated terms arising from
the determination of the propagator pole in \FH.


\section{$\DR$ parameters as input for an OS calculation}\label{FHwithDRinputSection}

In this section we discuss issues related to the conversion between
parameters of OS and $\DR$ renormalization schemes. While the discussion
will focus on the case where $\DR$ input parameters are converted into OS
ones that are then inserted into a result in the OS scheme, it should be
stressed that the related problems are not intrinsic to the OS approach. The
same problems would occur if a $\DR$ result were used with OS input 
parameters. The discussed problems are also not specific to Higgs mass
predictions in SUSY models, but would appear whenever there are numerically
large higher-order logarithms arising from a large splitting between the
relevant scales of the considered quantity.
In predictions for the mass of the SM-like Higgs boson within the MSSM,
the result is however particularly sensitive to higher-order effects of this
kind through the pronounced dependence on the 
stop mixing parameter $\Xt$, which receives large corrections when
converting from the $\DR$ to the OS scheme or vice versa. 

In the case where fixed-order results at the $n$-loop level obtained in two
different renormalization schemes are compared with each other, and
higher-order logarithms are unknown and not expected to be 
particularly enhanced, it is well known that the results based on the same
type of corrections in two schemes differ by terms that are of 
${\cal O}(n+1)$. The same is true for different options regarding how to
perform the parameter conversion that differ from each other by higher-order
contributions. The numerical differences observed in such a comparison 
can therefore be used as an indication of the possible size of unknown
higher-order corrections. 

The situation is different, however, in the case that we are considering
here, since the comparison is not performed between fixed-order results but
between results incorporating a series of (resummed) higher-order
logarithms. It is crucial in such a case that the correct form of the higher-order
logarithms that can be derived via EFT methods, which in our case arise 
from the large splitting between the assumed SUSY scale and the weak scale,
is maintained in the parameter conversion. We will demonstrate below that
the parameter conversion that is usually applied for a comparison of
renormalization schemes in fixed-order results does not maintain the correct
form of the higher-order logarithms. Since those higher-order logarithms are
numerically important, a conversion carried out in the described way leads
to very large numerical discrepancies for large values of the SUSY scale.


\subsection{Conversion between $\DR$ and OS parameters applicable to
fixed-order results}
\label{FHwithDRinputNaiveSection}

The most straightforward method used for the conversion of
$\DR$ input parameters to OS parameters in fixed-order results is to derive
the shift between a parameter $p$ in the two schemes according to
$p^\OS = p^\DR + \Delta p$ at the considered loop order, see e.g.\ 
\cite{Carena:2000dp}. Accordingly, at 
the full one-loop level, including logarithmic as well as
non-logarithmic terms, the conversion from $\DR$ to OS parameters 
for the stop mixing parameter and the stop masses, which are particularly
relevant in the context of MSSM Higgs mass predictions, reads
(for explicit formulas see
\cite{Degrassi:2001yf,Brignole:2001jy,Brignole:2002bz,Williams:2011bu})
\begin{align}
\Xt^\OS &= \Xt^\DR + \Delta \Xt, \label{XtConvEq}\\
\Mstope &= \mstope^\DR + \Delta \mstope, \\
\Mstopz &= \mstopz^\DR + \Delta \mstopz. \label{MSt2ConvEq}
\end{align}
Here $\Delta m_{\tilde t_{1,2}}$ is given by the corresponding
difference of the $\DR$ and the OS counterterm. In \FH, the shift
of $X_t$ is obtained by first calculating the OS stop masses and the
OS stop mixing angle $\theta_{\tilde t}^\OS$. These are then used to
obtain $X_t^\OS$ via  
\begin{align}
M_t X_t^\OS = (\Mstope^2-\Mstopz^2)\sin\theta_{\tilde t}^\OS\cos\theta_{\tilde t}^\OS.
\label{eq:XtOSFH}
\end{align}
Relating this prescription for $X_t^\OS$ to the 
$\DR$ input parameters $\Xt^\DR$, $\mstope^\DR$, $\mstopz^\DR$, one can see
that \Eq{eq:XtOSFH} contains products of one-loop contributions and
therefore involves higher-order terms. Alternatively one could have used an
expression for the conversion that is truncated at the one-loop level. The
difference between the two prescriptions would be of the order of unknown
higher-order corrections in a fixed-order comparison. 
The on-shell parameters obtained as described above are then used as input of the 
fixed-order OS renormalized calculation. This means in particular that the 
knowledge of the initial $\DR$ parameters is not used any further once the
conversion to OS parameters has been carried out. While this procedure is
suitable for fixed-order results, it leads to problems if results containing
a series of higher-order logarithms are meant to be converted.

Indeed, applying the described parameter conversion to the case of a
$\DR$ result that incorporates higher-order logarithms generates additional
higher-order terms causing a deviation in the logarithmic contributions.
This can be seen
by investigating the Higgs self-energy up to the two-loop level where the
parameter $\Xt^\OS$ obtained from the conversion has been inserted,
\begin{align}
\hat\Sigma_{hh}^\OS(\Xt^\OS) ={}& \hat\Sigma_{hh}^{(1),\OS}(\Xt^\OS) + \hat\Sigma_{hh}^{(2),\OS}(\Xt^\OS). 
\end{align}
Using instead \Eq{XtConvEq} to write $\Xt^\OS$ in terms of $\Xt^\DR$,
\begin{align}
\hat\Sigma_{hh}^\OS(\Xt^\OS) ={}& \hat\Sigma_{hh}^{(1),\OS}(\Xt^\DR + \Delta
\Xt) + \hat\Sigma_{hh}^{(2),\OS}(\Xt^\DR + \Delta \Xt),
\end{align}
and performing an 
expansion in $\Delta \Xt$ yields
\begin{align}
\hat\Sigma_{hh}^\OS(\Xt^\OS) ={}& \hat\Sigma_{hh}^{(1),\OS}(\Xt^\DR) + \left[\frac{\partial}{\partial \Xt}\hat\Sigma_{hh}^{(1),\OS}(\Xt^\DR)\right]\Delta \Xt \nonumber\\
&+ \hat\Sigma_{hh}^{(2),\OS}(\Xt^\DR) + \left[\frac{\partial}{\partial \Xt}\hat\Sigma_{hh}^{(2),\OS}(\Xt^\DR)\right]\Delta \Xt+ \order{\Delta \Xt^2} =\\
={}& \hat\Sigma_{hh}^\DR(\Xt^\DR) + \left[\frac{\partial}{\partial
\Xt}\hat\Sigma_{hh}^{(2),\OS}(\Xt^\DR)\right]\Delta \Xt + \order{\Delta\Xt^2}.\label{eq:Sihhconv}
\end{align}
Thus, the obtained expression obviously differs from the original $\DR$ result by
terms of 3-loop order and beyond. One would furthermore need 
to convert also all other parameters entering the self-energy to the $\DR$ scheme 
in order to exactly
recover the $\DR$ renormalized self-energy. 


\subsection{The case of large higher-order logarithms}
\label{FHwithDRinputFOSection}

The higher-order terms in
\Eq{eq:Sihhconv} that are not present in the original $\DR$ result contain
in general logarithmic contributions which for a result containing a series
of higher-order logarithms cause a deviation from the
logarithmic corrections determined via the RGE. In our numerical discussion
in \Sec{NumericalResultsSection} below we will demonstrate that those
higher-order contributions that are induced by the parameter conversion are
indeed numerically sizeable.

Another issue that is relevant in a hybrid approach,
as pursued in \FH, where a fixed-order
result in the OS scheme is combined with higher-order logarithmic
expressions that are expressed in the $\DR$ scheme 
concerns the $\DR$ value of $\Xt$ that is 
used in the EFT part of the calculation. Only logarithmic terms are kept in the 
relation between $\Xt^{\DR,\EFT}$ and $\Xt^\OS$, see \Eq{XtEFT}. 
If instead an input value for $\Xt^\DR$ were converted to $\Xt^\OS$ using
the full one-loop contributions according to \Eq{XtConvEq}, the 
stop mixing parameter used in the EFT calculation of \FH, $\Xt^{\DR,\EFT}$, 
would differ from the input parameter $\Xt^\DR$.

In order to properly address the case where $\DR$ parameters
associated with a result containing a series of higher-order logarithms 
are used as input for \FH, we follow the strategy to perform the parameter
conversion in the fixed-order result rather than in the infinite series of
higher-order logarithms. For this purpose we have extended \FH\ 
such that the incorporated fixed-order result is given in terms of the 
$\DR$ parameters $\Xt^\DR$, $\mstope^\DR$, $\mstopz^\DR$ 
(the actual input parameters are the soft-breaking parameters of the stop sector).
This new result
complements the existing result that is given in terms of the 
on-shell parameters $X_t^\OS$, 
$\Mstope \equiv \mstope^\OS$, $\Mstopz \equiv \mstopz^\OS$.
The reparametrisation on which the new result is based can be viewed as the
parameter conversion described in the example of the previous section, 
but truncated at the two-loop level,
\begin{align}\label{FOconvTLterm}
\hat\Sigma_{hh}^\OS(\Xt^\OS)\to{}& \hat\Sigma_{hh}^{\OS}(\Xt^\DR) +
\left[\frac{\partial}{\partial
\Xt}\hat\Sigma_{hh}^{(1),\OS}(\Xt^\DR)\right]\Delta \Xt =\hat\Sigma_{hh}^\DR(\Xt^\DR). 
\end{align}
We have used the same procedure as the one described here 
for the stop mixing parameter also for the stop masses. 
The two-loop terms that are induced by the conversion at the one-loop
level have been added to the two-loop result derived in the on-shell scheme
in order to arrive at the corresponding expression in the $\DR$ scheme.
Compact expressions for these additional terms valid in the case $\msusy\gg M_t$ and degenerate $M_L=M_{\tilde t_R}=\msusy$ can be found in
\App{TLaddedterms}. 
It should be noted
that we would have obtained the same result if we had performed the 
diagrammatic calculation with a $\DR$ renormalization of the respective
parameters 
instead of reparametrizing the final result.
Using the above result given in terms of $\DR$ parameters, the 
value of $\Xt$ that is used in the EFT part of the calculation equals the
$\DR$ input parameter,
$\Xt^{\DR,\EFT} = \Xt^\DR$.
For this setting in \FH\ with
$\DR$ input parameters 
the subtraction terms have been adjusted 
such that the logarithms already contained
in the fixed-order result for the 
$\DR$ renormalized self-energy are subtracted (rather than the ones contained 
in the OS renormalized self-energy, as it is the case for OS input
parameters).

Accordingly, depending on the provided input parameters the evaluation of
the prediction for the mass of the SM-like Higgs boson in \FH\ proceeds in
the following ways:
\begin{itemize}

\item
For on-shell input parameters the on-shell fixed-order result is combined
with the higher-order logs obtained in the EFT approach, where 
$\Xt^{\DR,\EFT}$ is related to $\Xt^\OS$ as specified in \Eq{XtEFT}. 

\item
For $\DR$ input parameters in the stop sector 
associated with a result containing a series of higher-order logarithms 
the $\DR$ fixed-order result is combined
with the higher-order logs obtained in the EFT approach, where 
$\Xt^{\DR,\EFT} = \Xt^\DR$.

\item
For $\DR$ input parameters in a low-scale SUSY scenario where the impact of
higher-order logarithms is expected to be small, both the fixed-order $\DR$
result and the fixed-order on-shell result can be employed, where for the
latter the parameter conversion described in the previous section is used.

\end{itemize}


\section{Comparison of \FH to other codes}\label{EFTcompSection}

In the previous sections, we investigated methodical
differences between the different approaches
for predicting the lightest 
\cp-even Higgs boson mass in
the MSSM, focusing in particular on the
comparison of the hybrid approach implemented in \FH with a pure EFT
calculation. In the following, we compare \FH numerically to other
codes. 

Publicly available codes based on diagrammatic fixed-order results or effective potential methods include {\tt CPSuperH}
\cite{Lee:2003nta,Ellis:2006eh,Lee:2007gn}, {\tt SoftSUSY}
\cite{Allanach:2001kg}, {\tt SPheno} \cite{Porod:2003um,Porod:2011nf}
and {\tt SUSPECT} \cite{Djouadi:2002ze}. Publicly available pure EFT
calculations are {\tt SUSY\-HD} \cite{Vega:2015fna} and {\tt MhEFT}
\cite{Draper:2013oza,Lee:2015uza,Gabe:2016}. \FS
 \cite{Athron:2014yba}, based on {\tt SARAH}
\cite{Staub:2009bi,Staub:2010jh,Staub:2012pb,Staub:2013tta}, includes
both a diagrammatic and an EFT result. Furthermore, it also has
the option to use a hybrid method different from the one pursued in \FH,
called \FEFT \cite{Athron:2016fuq}. 
Its basic idea is to include terms suppressed by the SUSY
scale into the matching conditions in order to obtain accurate results for both low and high scales. Recently, the same approach has been 
included into {\tt SPheno} \cite{Staub:2017jnp}.

The different levels of higher-order corrections implemented in the
various diagrammatic codes are listed in \cite{Draper:2016pys}. A
detailed numerical comparison between various diagrammatic and EFT codes
can be found in \cite{Athron:2016fuq}. In there, it is also discussed in
detail how \FEFT compares to other codes. We therefore focus in this
work on a comparison of \FH to \SHD as an exemplary EFT calculation.  

Before we can investigate the impact of the effects discussed in the
previous Sections on the comparison of {\tt Feyn}\-{\tt Higgs} and \SHD, we have to ensure
that the RGE results, i.e.\ the results for $\lambda(M_t)$, of both codes
are compatible with each other. Both codes implement full leading and
next-to-leading resummation and \order{\als\alt,\alt^2}
next-to-next-to-leading resummation of large logarithms. So the levels
of accuracy are basically identical. There are however several
differences which are listed below. 
\begin{itemize}
\item 
\SHD by default uses the top-Yukawa coupling extracted at the NNNLO
level. \FH instead uses the NNLO value by default, which is formally 
the appropriate setting for the resummation of
NNLL contributions. For
all numerical results shown in this work, we deactivate the NNNLO
corrections to the top-Yukawa coupling in \SHD. 
\item 
\SHD includes the bottom- and tau-Yukawa couplings in the
renormalization group running and also includes corresponding one-loop
threshold corrections. In {\tt Feyn}\-{\tt Higgs}, the bottom and tau Yukawa couplings are
set to zero in the EFT calculation. In the fixed-order diagrammatic
calculation, however, terms proportional to the bottom Yukawa coupling
are included at the one- and two-loop level (at the one-loop level for the
case of the tau Yukawa coupling). 
\item 
\SHD includes the electroweak gauge couplings in the running up to the three-loop level. \FH takes them into account up to the two-loop level. At the three-loop level, they are set to zero.
\item
\FH includes a one-loop running of $\tb$ to relate
$\tb(M_t)$, which is used as input of \FH, to $\tb(\msusy)$,
which enters through the matching at the SUSY scale. 
In contrast, \SHD uses $\tb(\msusy)$ as input. 
\item
Similarly, \FH uses a $\DR$ renormalized Higgsino mass parameter $\mu$ at the scale $M_t$. The running to the scale $\msusy$, at which it enters the EFT calculation via the matching conditions at the SUSY scale, is neglected. \SHD uses $\mu(\msusy)$ as input.
\end{itemize}
More details on the implemented EFT calculations are given in
\cite{Vega:2015fna,Bahl:2016brp}.  

Despite the listed differences including the different treatment of the renormalization scales of $\tb$ and $\mu$, we find excellent agreement between the
results of the RGE running of both codes. The numerical difference
of the quantity $v^2\lambda(M_t)$ calculated using the two codes is
always $\lesssim 50$ GeV$^2$ for the single scale scenario defined in
\Eq{MSusyDef} and $\tb \sim \order{10}$. This translates into a
difference in $M_h$ of~$\lesssim 0.1$~GeV.


\section{Numerical results}\label{NumericalResultsSection}

In this Section, we present a numerical investigation of the effects
discussed in the previous Sections and compare the result obtained by
\FH to \SHD as an exemplary pure EFT code. We restrict ourselves
to the single scale scenario defined
in \Eq{MSusyDef}. Apart from the parameters of the stop sector, we neglect all renormalization scheme conversions necessary to relate the parameters of \Eq{MSusyDef} as defined in \FH to the parameters as defined in \SHD. We furthermore set 
\begin{align}
\tb = 10,
\end{align}
i.e. we use $\tb(M_t)=10$ as input for \FH and $\tb(\msusy)=10$ as
input for \SHD. As mentioned in \Sec{EFTcompSection}, the difference in the renormalization scales is negligible for the considered scenario. All soft-breaking trilinear couplings except the one of the scalar top
quarks are choosen to be
\begin{align}
A_{e,\mu,\tau,u,d,c,s,b} = 0.
\end{align}
For all soft-breaking parameters (i.e. those of the stop sector), we use the $\DR$ scheme with the renormalization scale being $\msusy$.
If not stated otherwise, we use a parametrization of the non-logarithmic
contributions in terms of the SM $\MS$ NNLO top mass and $v_{G_F}$
(see \Sec{NonLogCompSection}), corresponding to
choosing \mbox{{\tt runningMT = 1}} as \FH flag. 

\begin{figure*}\centering
\begin{minipage}{.48\textwidth}\centering
\includegraphics[width=\textwidth]{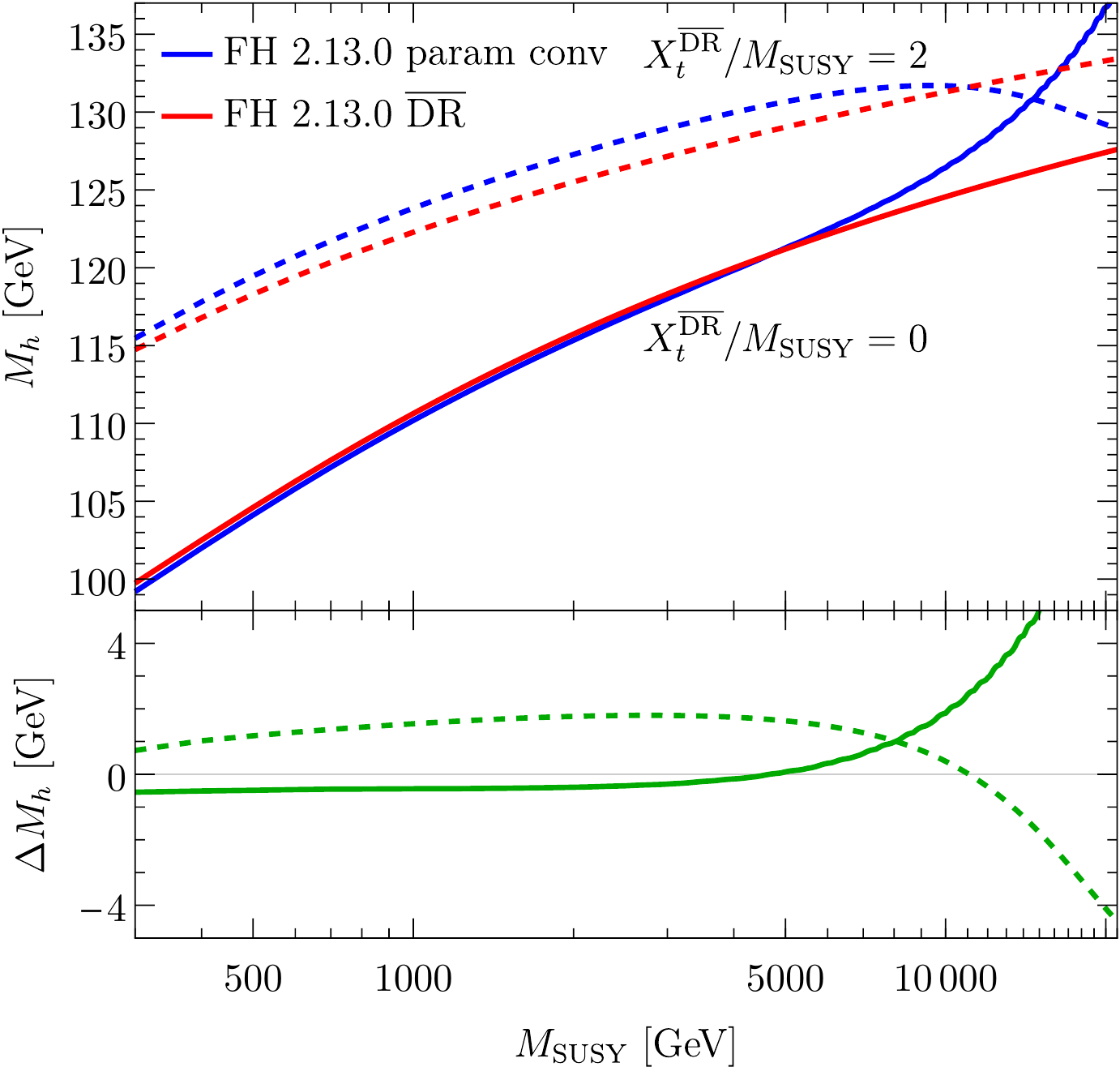}
\end{minipage}
\hfill
\begin{minipage}{.48\textwidth}\centering
\includegraphics[width=\textwidth]{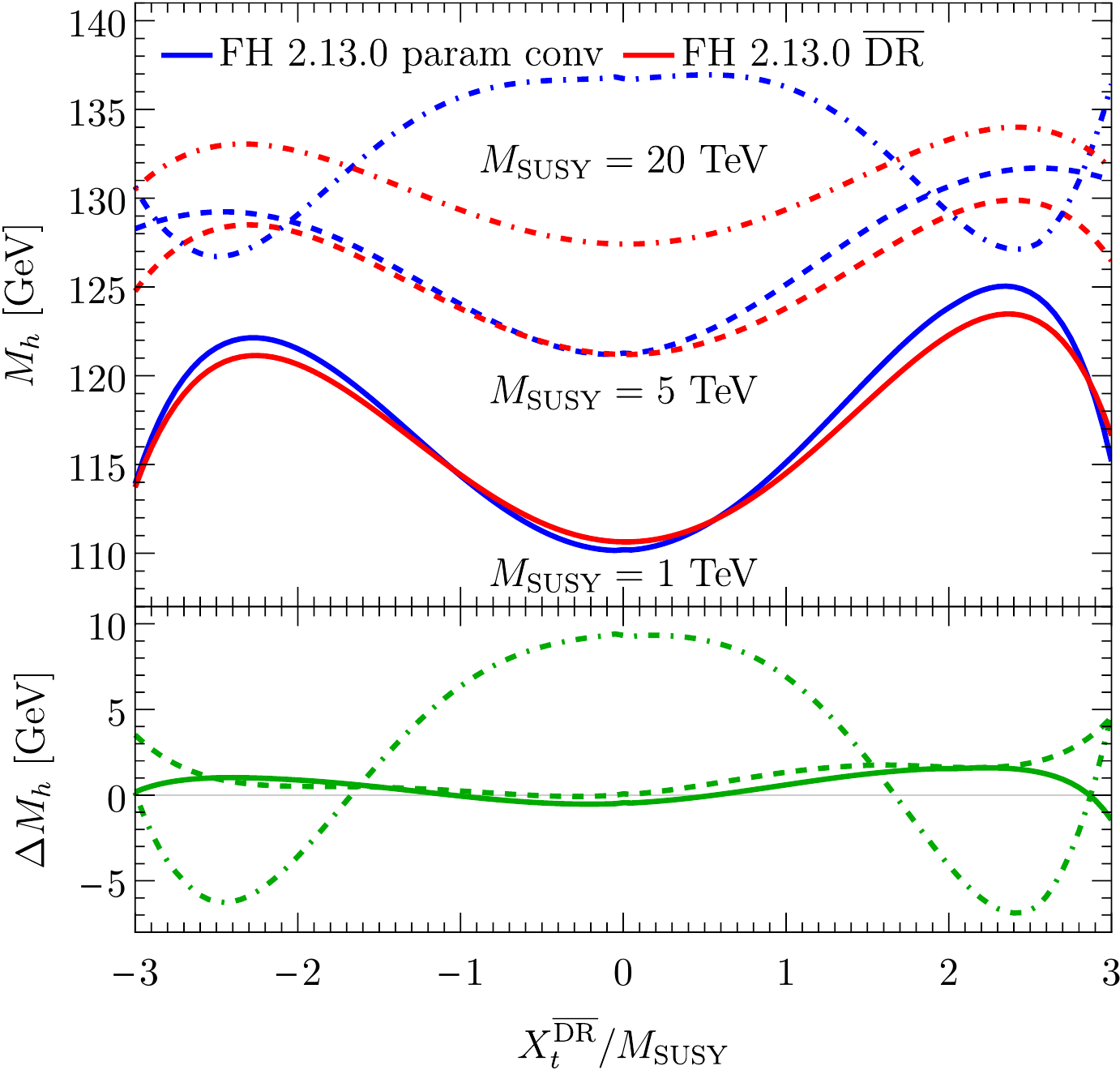}
\end{minipage}
\caption{Left: $M_h$ as a function of $\msusy$ for $\Xt^\DR/\msusy=0$
  (solid) and $\Xt^\DR/\msusy=2$ (dashed). The results of {\tt FeynHiggs2.13.0} with a $\DR$ to OS conversion of the input parameters (blue) and a $\DR$ renormalization of the fixed-order result (red) are compared. Right: Same as left plot, apart that $M_h$ is
  shown in dependence of $\Xt^\DR/\msusy$ for $\msusy = 1$ TeV
  (solid), $\msusy = 5$ TeV (dashed) and $\msusy = 20$ TeV
  (dot-dashed). In the bottom panels, the difference between the blue and
red curves is shown ($\Delta M_h = M_h(\text{FH 2.13.0 param conv}) -
M_h(\text{FH 2.13.0 $\DR$})$).} 
\label{FigConvComp}
\end{figure*}

We first look at the numerical difference between employing the type of conversion
from $\DR$ to OS input parameters which is suitable for the comparison of
fixed-order results (``FH 2.13.0 param conv'') 
and using a $\DR$ renormalized fixed-order result (``FH 2.13.0 $\DR$''), see
the discussion in
\Sec{FHwithDRinputSection}. The left plot of \Fig{FigConvComp} shows
the corresponding results
for $\Xt^\DR/\msusy = 0\;(2)$ as solid (dashed)
lines as a function of $\msusy$. One can see
that for $\msusy\lesssim 5$ TeV the difference between the two methods
leads to an approximately constant shift in the prediction for
$M_h$. For vanishing mixing the prediction obtained by using a $\DR$
renormalized fixed-order result is $\sim 0.5$ GeV higher
than the one obtained by a naive
scheme conversion of the input parameters. For $\Xt/\msusy = 2$,
the shift is larger. The result
obtained using a $\DR$ fixed-order result is $\sim 1- 1.5 \gev$
smaller than the 
one obtained by the naive
conversion of the input parameters. The shifts occur
not only for scales
of a few TeV, but also for very low scales ($\msusy\simeq 0.3 \tev$).
Therefore, we conclude that at low scales the observed shifts are
mainly caused by non-logarithmic 
higher-order terms 
by which the $\DR$ result and the result involving a parameter conversion
differ from each other.
As usual, non-logarithmic terms tend to be
larger for $|\Xt^\DR/\msusy|\sim 2$ than for vanishing stop mixing. 

For $\msusy \gtrsim 5$ TeV, we observe that the difference be-tween the
two results is increasing rapidly to up to $10 \gev$ for vanishing mixing
and up to $5 \gev$ for $|\Xt^\DR/\msusy|\sim 2$ in the region up to
$\msusy \sim 20 \tev$. This behavior is mainly due to
the fact that the parameter conversion that is used for the comparison of
fixed-order results induces higher-order logarithmic contributions that are
not compatible with the implemented resummation of logarithms to all orders 
(see the discussion in \Sec{FHwithDRinputNaiveSection}). 
For high SUSY scales, where the higher-order logarithmic contributions
become numerically large, this mismatch leads to the observed large
deviations. To a lesser extent, also the deviation between 
the input $\Xt^\DR$ and the $\Xt^{\DR,\EFT}$ used in
the EFT calculation plays a role in this context, see
\Sec{FHwithDRinputFOSection}.

In the right plot of \Fig{FigConvComp} the two results
are compared as a function of $\Xt^\DR/\msusy$ for $\msusy = 1, 5, 20 \tev$,
shown as solid, dashed and dot-dashed lines, respectively.
For $\msusy = 1 \tev$ and $\msusy = 5 \tev$ 
the deviations stay relatively small except for
the highest values of $|\Xt^\DR/\msusy|$. In contrast, for $\msusy = 20 \tev$ 
the uncontrolled higher-order contributions induced by the 
naive conversion of the input parameters are seen to have a huge effect
which even reverts the usual pattern of the dependence on 
$|\Xt^\DR/\msusy|$, giving rise to local minima at 
$|\Xt^\DR/\msusy|\simeq \pm 2.3$. 
We emphasize again that the same kind of uncontrolled higher-order effects
would occur if a naive conversion of OS to $\DR$ parameters would be used as
input for a $\DR$ result containing a series of numerically large
higher-order logarithms.
\Fig{FigConvComp} shows that numerical instabilities noticed in comparisons
of EFT results with \FH carried out in the literature are a consequence of
an inappropriate application of the conversion of input parameters between 
the OS and the $\DR$ schemes. The higher-order contributions implemented in 
\FH are seen to be numerically stable up to very high SUSY scales in the
considered scenario.

\begin{figure*}\centering
\begin{minipage}{.48\textwidth}\centering
\includegraphics[width=\textwidth]{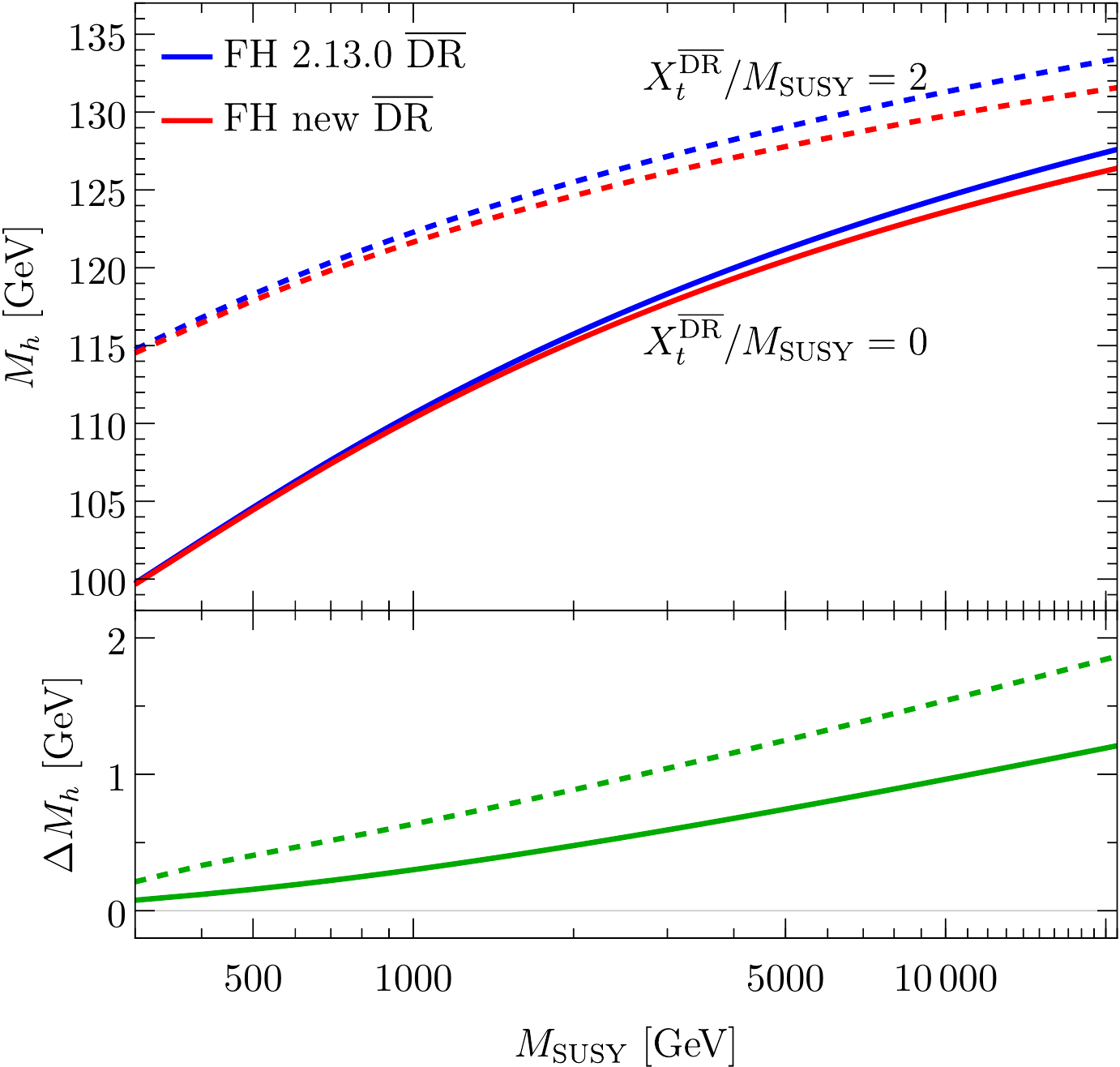}
\end{minipage}
\hfill
\begin{minipage}{.48\textwidth}\centering
\includegraphics[width=\textwidth]{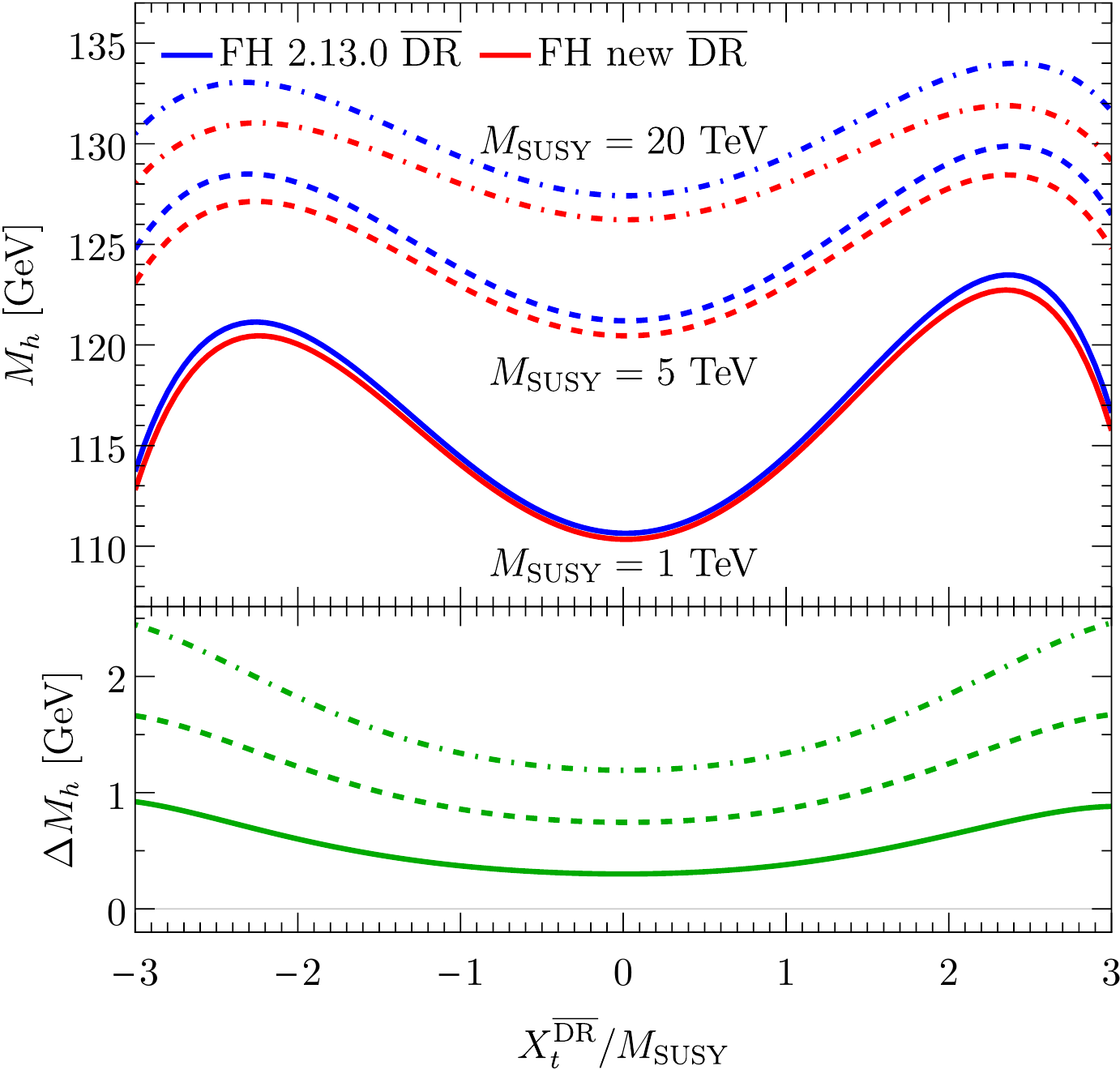}
\end{minipage}
\caption{Comparison of the $M_h$ predictions of
{\tt FeynHiggs2.13.0 $\DR$} with {\tt FeynHiggsnew $\DR$}, where in the
new version terms arising from the determination of the propagator pole
are omitted that go beyond the level of the corrections implemented in the
irreducible self-energies.
Left: 
Prediction for $M_h$ as function of
$\msusy$ for vanishing stop mixing and $\Xt^\DR/\msusy = 2$.
Right: 
Prediction for $M_h$ as function of
of $\Xt^\DR/\msusy$ for $\msusy = 1$
  TeV (solid), $\msusy = 5$ TeV (dashed) and $\msusy = 20$ TeV
  (dot-dashed). In the bottom panels, the difference between the blue and red curves is shown ($\Delta M_h = M_h(\text{FH 2.13.0 $\DR$})-M_h(\text{FH new $\DR$})$).} 
\label{Figp2Dep}
\end{figure*}

For the further \FH results shown below we use the $\DR$ renormalization of
the parameters in the stop sector. As a next step 
we investigate the impact of the terms arising from the determination of 
the propagator pole. As explained in \Sec{CompSection}, there occurs a
cancellation in the limit of a large SUSY scale between 
non-SM terms arising through the determination of the propagator pole and
contributions from the subloop renormalization of the irreducible
self-energy diagrams. While up to the version 
{\tt FeynHiggs2.13.0} this cancellation was incomplete for terms beyond
\order{\alt^2,\alt\alb,\alb^2} (see \Eq{FHpoleMass}), we have modified the
determination of the propagator poles in the new version of \FH such that
terms are omitted that would not cancel because their counterpart in the
irreducible self-energies is not incorporated at present. 
In \Fig{Figp2Dep} {\tt FeynHiggs2.13.0 $\DR$} is compared with the new
version, which is labelled as {\tt FeynHiggsnew $\DR$}. The difference
between the two results corresponds essentially
to the terms $\Delta_{p^2}^\text{logs}$ and $\Delta_{p^2}^\text{nolog}$ given
in \Eqs{p2TermsLog}{p2TermsNonLog}.
In the left plot of \Fig{Figp2Dep}, we show the results as a
function of $\msusy$ for $\Xt^\DR = 0$ and $\Xt^\DR/\msusy = 2$. One observes that the difference grows nearly
logarithmically with $\msusy$. This is expected since the largest terms
in $\Delta_{p^2}^\text{logs} + \Delta_{p^2}^\text{nolog}$ are in fact
logarithms of the SUSY scale over $M_t$. Consequently, for small scales ($\msusy \lesssim 1$ TeV), these
terms induce only a small upwards shift of $\lesssim 0.5 \gev$. For large
scales however ($\msusy\gtrsim 5 \tev$), this shift grows to up to $1.5 \gev$ for vanishing stop mixing and $2 \gev$ for $\Xt^\DR/\msusy = 2$.
In the right plot of \Fig{Figp2Dep} the difference is depicted as a
function of $\Xt^\DR/\msusy$ for $\msusy = 1, 5, 20 \tev$, shown as
solid, dashed and dot-dashed lines, respectively. One can
see that the difference between the two results
is approximately quadratically depependent on
$\Xt^\DR/\msusy$. This reflects the $X_t^\DR$ dependence of the derivative
of the Higgs-boson self-energy (see \Eq{dsehh_stop_Eq} below).

\begin{figure*}\centering
\begin{minipage}{.48\textwidth}\centering
\includegraphics[width=\textwidth]{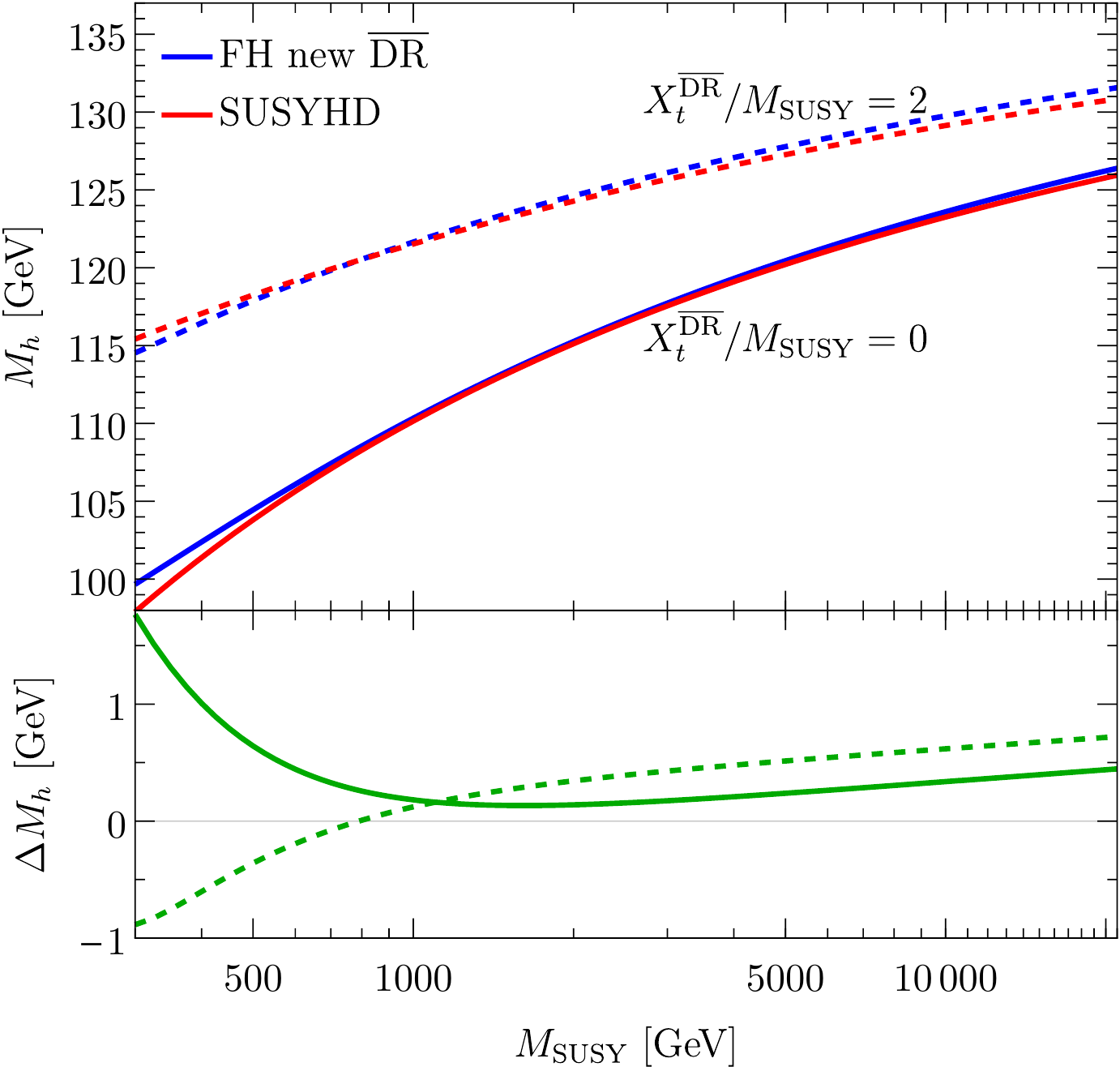}
\end{minipage}
\hfill
\begin{minipage}{.48\textwidth}\centering
\includegraphics[width=\textwidth]{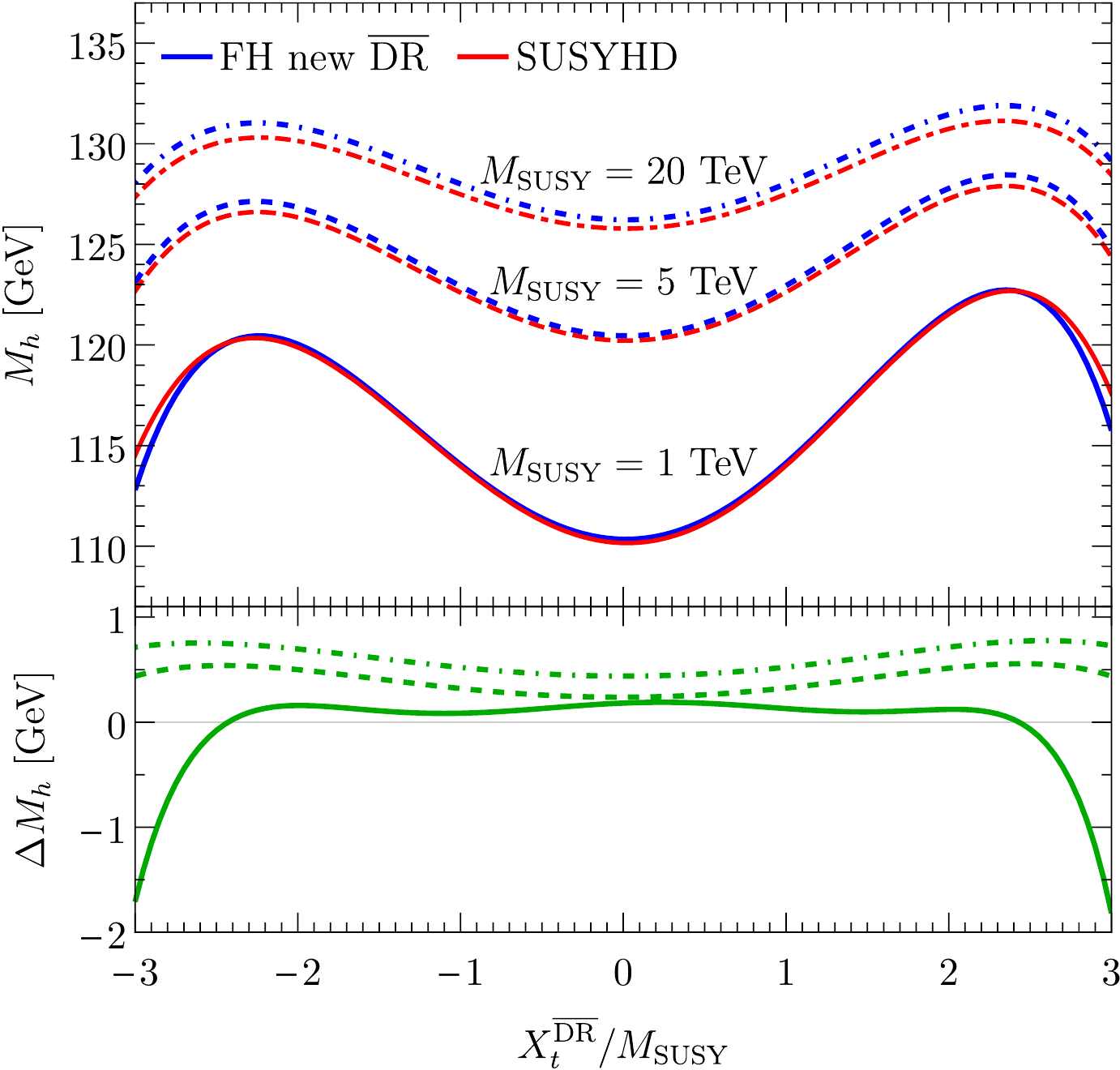}
\end{minipage}
\caption{Comparison of the $M_h$ predictions of
{\tt FeynHiggsnew $\DR$} with \SHD.
Left: $M_h$ as function of $\msusy$ for $\Xt^\DR/\msusy=0$
(solid) and $\Xt^\DR/\msusy=2$ (dashed). 
Right: $M_h$ as function of
$\Xt^\DR/\msusy^\DR$ for $\msusy = 1$ TeV (solid), $\msusy = 5$ TeV (dashed)
and $\msusy = 20$ TeV (dot-dashed). 
In the bottom panels, the difference
between the blue and red curves is shown ($\Delta M_h = M_h(\text{FH new
$\DR$})-M_h(\text{SUSYHD})$).}
\label{FigFHvsSHD}
\end{figure*}

Having investigated the numerical impact of the scheme conversion of the
input parameters as well as of the terms arising from the determination of
the propagator pole, we now turn to a direct
comparison of \FH with \SHD.%
\footnote{
We remind the reader that we use \SHD with the top Yukawa coupling
evaluated at the NNLO level. Using instead the NNNLO value would shift the results of \SHD shown here downwards by $\sim 0.5\gev$.
}
The \FH results in this comparison are the ones of the new version,
{\tt FeynHiggsnew $\DR$}, where the stop sector is renormalized in the $\DR$
scheme and terms arising from the determination of the propagator pole
are omitted that go beyond the level of the corrections implemented in the
irreducible self-energies, as described above.

The left plot of \Fig{FigFHvsSHD} shows $M_h$
as a function of $\msusy$ for $\Xt^\DR/\msusy = 0\;(2)$ as solid
(dashed) lines. 
For vanishing stop mixing and $\msusy\gtrsim 1 \tev$, we observe an
excellent agreement of the \FH curve with the \SHD result. Even for
very large scales $\msusy\simeq 20 \tev$, we find agreement within 
$\sim 0.5 \gev$ in the considered simple numerical scenario, in which all
SUSY scales are chosen to be equal to each other. 
For low scales ($\msusy\lesssim 1 \gev$), it can be seen 
that the \FH result is higher by up to $\sim 1.7 \gev$ compared to
the \SHD result. The origin of this difference are terms suppressed by
the SUSY scale, which are included in \FH but not in \SHD, as will
be discussed below.
For $\Xt^\DR/\msusy=2$, we basically observe the same behavior
as in case of vanishing stop mixing. The overall agreement in the
simple numerical scenario is very good
(within $\sim 0.7 \gev$ for $\msusy\gtrsim 0.5 \tev$). For low scales
($\msusy\lesssim 0.5 \gev$), the \FH result is
lower compared to the \SHD result by up to $\sim 1 \gev$.
As in the case of vanishing stop mixing, this can be traced back to terms
suppressed by the SUSY scale. We will discuss this and investigate 
the remaining differences in more detail below.

In the right plot of \Fig{FigFHvsSHD} the comparison between the $M_h$
prediction of the new \FH version and \SHD is shown 
as a function of $\Xt^\DR/\msusy$ for
$\msusy = 1, 5, 20 \tev$, shown as solid, dashed and dot-dashed lines,
respectively. Again one can
see an overall very good agreement between both
codes for $\msusy\gtrsim 1\tev$ (within $1 \gev$) in the considered
simple numerical scenario. 
The agreement is especially good for small
$|\Xt^\DR/\msusy|$, but the deviations stay below $1 \gev$ also for
increasing mixing in the stop sector except for the highest values of 
$|\Xt^\DR/\msusy|$ in the case of $\msusy = 1\tev$.
The larger deviations of up to $\sim 2\gev$ for
$|X_t^\DR/\msusy|\gtrsim 2.5$ in the case of $\msusy = 1\tev$ are
due to terms suppressed by $\msusy$
which become large for increasing $|\Xt^\DR/\msusy|$.

\begin{figure*}\centering
\begin{minipage}{.48\textwidth}\centering
\includegraphics[width=\textwidth]{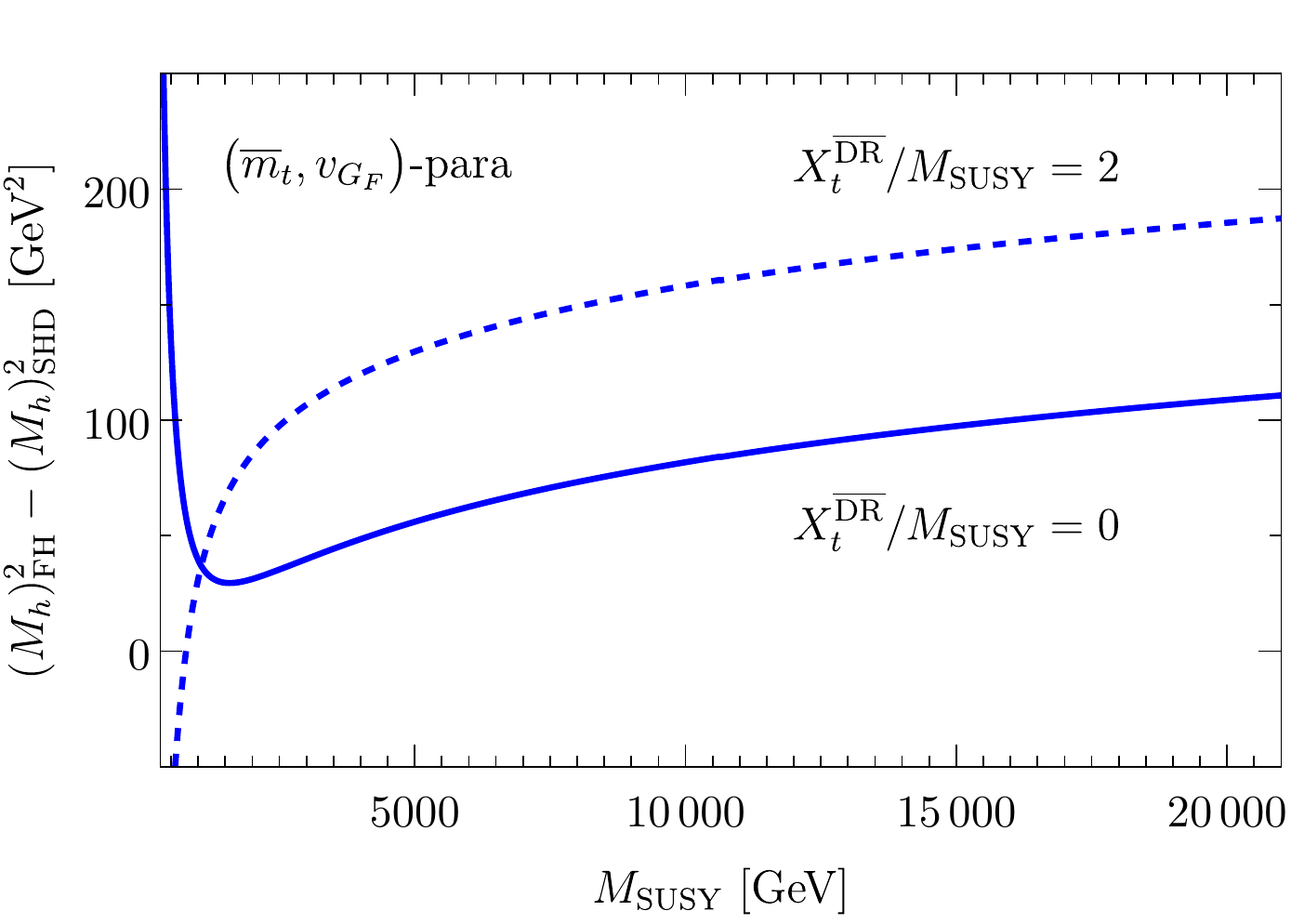}
\end{minipage}
\hfill
\begin{minipage}{.48\textwidth}\centering
\includegraphics[width=\textwidth]{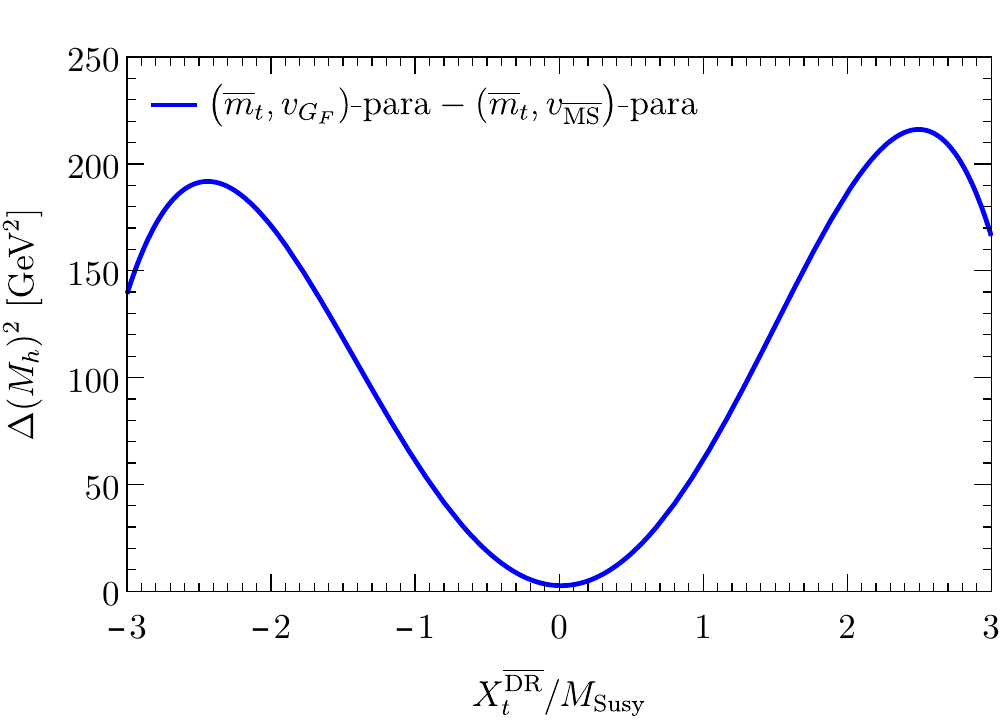}
\end{minipage}
\caption{Left: Difference of the $M_h^2$ predictions of
{\tt FeynHiggsnew $\DR$} and \SHD as a
  function of $\msusy$ for $\Xt^\DR/\msusy=0$ (solid) and
  $\Xt^\DR/\msusy=2$ (dashed). For the parametrization of the
  diagrammatic result of \FH the SM NNLO $\MS$ top-quark mass is chosen.
  Right: Differences due to the different parametrization of the top-quark
mass and the vev in a fixed-order 
\order{\alt\als,\alt^2}
  calculation, 
  taking into account only non-logarithmic terms, as a function of
  $\Xt^\DR/\msusy$. The difference between 
  the result parametrized in terms of the $\MS$ NNLO top-quark mass and
$v_{G_F}$ and the one parametrized in terms of the $\MS$ NNLO top-quark mass
and $v_\MS$ is shown.
} 
\label{FigFHvsSHDsq}
\end{figure*}

In \Fig{FigFHvsSHDsq}, we further investigate these remaining differences
between \FH and \SHD observed in \Fig{FigFHvsSHD}. In the left plot 
we show the difference between the
results of \FH and \SHD for $M_h^2$ (not for $M_h$). Since in both
codes actually $M_h^2$ is calculated, taking the square root of
these results can obscure the true dependences of the difference. 
As an example, if
the difference in $M_h^2$ is constant when varying $\msusy$, we would
not observe a constant difference when comparing the difference in
$M_h$. We show in the plot the difference in
$M_h^2$ for the case where
the fixed-order result of \FH is parametrized in terms of the SM NNLO $\MS$
top mass. For $\msusy\lesssim 1 \tev$ in the case of vanishing mixing
and for $\msusy\lesssim 3 \tev$ in the case of
$X_t^\DR/\msusy=2$ we observe large gradients.  
For larger scales
($\msusy\gtrsim 3 \tev$), the difference is only slowly increasing when
raising $\msusy$. For vanishing stop mixing, the difference is growing
by $\sim 50 \gev^2$ when raising $\msusy$ from 3 TeV to 20 TeV. For
$X_t^\DR/\msusy=2$, similarly a growth of $\sim 50 \gev^2$ is
recognizable. This behavior is mostly due to the differences in the EFT
calculations implemented in \FH and \SHD discussed 
in \Sec{EFTcompSection}. In addition, however, we observe an offset relative
to the zero axis for $\msusy\gtrsim 3$~TeV. 
For vanishing stop mixing, it is small ($\sim 50 \gev^2$),
whereas for $X_t^\DR/\msusy=2$, the shift is more significant ($\sim
150$~GeV$^2$).  
The nearly constant offset 
between the two codes can be traced
back to the
different parametrization of the non-logarithmic terms discussed in
\Sec{NonLogCompSection}. 

We further analyse the influence of
the different ways to parametrize the non-logarithmic terms in the right
plot of \Fig{FigFHvsSHDsq}. It shows the difference in $M_h^2$ obtained from
a diagrammatic calculation of \order{\alt\als,\alt^2} using
different parametrizations of the 
vev for the
non-logarithmic one- and two-loop terms (see \Sec{NonLogCompSection} for more details). Note that these non-logarithmic terms, apart of \order{v/\msusy} contributions, stay constant when varying $\msusy$. 
For $\Xt^\DR/\msusy \sim 2$ the difference between 
parametrizations in terms of 
$v_{G_F}$ and $v_\MS$
(both using the SM NNLO $\MS$ top-quark mass)
amounts to $\sim 170 \gev^2$. Such a shift accounts for the main part of the nearly constant offset
observed in the left plot of
\Fig{FigFHvsSHDsq}. 
For $\Xt^\DR/\msusy \sim 0$ the difference between the
parametrizations in terms of $v_{G_F}$ and $v_\MS$ is seen to become
very small.
The nearly constant offset for vanishing stop mixing observed 
in the left plot of \Fig{FigFHvsSHDsq}
can be explained in a similar way by different parameterization of
terms that are not of \order{\alt\als,\alt^2}.

Finally, we briefly comment on the differences between \FH and other codes
that have been reported in the literature. In \cite{Vega:2015fna} it was claimed that
differences between \FH and \SHD of up to $\sim 9$ GeV would occur
for $\msusy = 2\tev$ and $X_t^\DR/\msusy \sim \sqrt{6}$. As already noted in
\cite{Vega:2015fna}, this difference was somewhat reduced if the NNLO
$\MS$ top mass was employed in the calculation of \FH.%
\footnote{In the \FH version employed in the comparison 
by default the NLO $\MS$ top mass was used. This was formally
correct for the resummation of the LL and NLL contributions that was
implemented in \FH at that time. Numerically, 
the shift in the top-quark mass from NLO to NNLO generated the main effect 
when going to NNLL resummation~\cite{Bahl:2016brp}.} 
While at the time of the comparison carried out in 
\cite{Vega:2015fna} 
the EFT calculation of \FH was not yet at the same level
of accuracy as the one of \SHD, the differences claimed in \cite{Vega:2015fna} were in fact primarily
caused by 
an inappropriate application of the conversion of input parameters between
the $\DR$ and the OS scheme. The inappropriate parameter conversion, for
which the authors of \cite{Vega:2015fna} used their own routine, caused a
deviation
of 3--$4\gev$ for $\msusy = 2\tev$ and $X_t^\DR/\msusy \sim \sqrt{6}$ and
was also responsible for the apparent numerical instability 
at large SUSY scales of the \FH curve
with $X_t^\DR/\msusy = 0$ shown in \cite{Vega:2015fna}. The numerical effect
of this deviation was larger than the shift caused by employing 
the NNLO or NNNLO $\MS$ top-quark mass in \FH, in contrast to the claim made
in \cite{Vega:2015fna}.

Also the comparison figures shown in
\cite{Athron:2016fuq,Staub:2017jnp} are plagued by deficiencies arising from an 
inappropriate application of the parameter conversion between
the $\DR$ and the OS scheme. 
We stress again that such a parameter conversion would give rise to the same
kind of problems when starting from OS parameters and converting to 
$\DR$ ones.


\section{Conclusions}\label{ConclusionsSec}

We have presented a detailed comparison between various approaches used
to predict the mass of the SM-like Higgs boson in the MSSM in a scenario in
which all SUSY mass scales are chosen equal to each other. In particular we have
compared pure EFT calculations with the hybrid approach, in which
an explicit Feynman-diagrammatic fixed-order result is combined with the
leading higher-order contributions obtained from EFT methods.
In the literature significant deviations between the results obtained via
the two approaches have been reported especially at large SUSY scales.
In this work, we have identified three sources of the
observed differences. 

We could show that a large part of the reported discrepancies can be traced
back to parameter conversions between different renormalization schemes. 
In EFT
calculations typically the $\DR$ scheme is used for the renormalization
of SUSY breaking parameters, e.g.\ the stop mixing parameter. In the
diagrammatic calculation of \FH (in the default case) however, the OS scheme is employed
in the scalar top sector. We have demonstrated that the usual scheme
conversion of input parameters used for the comparison of fixed-order
results is not suitable for the comparison of results
containing a series of higher-order logarithms. This kind of parameter
conversion would induce higher-order
logarithmic contributions that are not compatible with the implemented
resummation of logarithms to all orders. We have shown that
the form of the higher-order logarithms obtained in one scheme can
manifestly be maintained if the fixed-order part of the calculation is
consistently reparametrized to this scheme. In order to enable this
approach for $\DR$ input parameters, we have extended \FH\ such that the
results are provided both in terms of the on-shell parameters $X_t^\OS$, 
$\Mstope \equiv \mstope^\OS$, $\Mstopz \equiv \mstopz^\OS$ (as before) 
and the $\DR$ parameters $\Xt^\DR$, $\mstope^\DR$, $\mstopz^\DR$. In practice,
this was achieved by reparametrizing the existing OS fixed-order result. 
We have demonstrated that many of the apparent discrepancies reported in the
literature have mainly been caused by 
an inappropriate application of the conversion of input parameters between
the OS and the $\DR$ schemes. It should be emphasized 
that this issue is not a problem of the OS
renormalization, but analogously appears if OS parameters are used as
input for codes employing the $\DR$ scheme.  

Another difference between pure EFT calculations and the hybrid approach
arises from the determination of the poles of the Higgs propagator matrix. We have shown explicitly at the two-loop level that
there occurs a
cancellation in the limit of a large SUSY scale between
non-SM terms arising through the determination of the propagator pole and
contributions from the subloop renormalization of the irreducible
self-energy diagrams. Since we expect that similar cancellations will happen at higher loops,
we have modified the
determination of the propagator poles in the new version of \FH such that
terms are omitted that would not cancel because their counterpart in the
irreducible self-energies is not incorporated at present. Unless otherwise stated, the numerical results presented in this paper have
been obtained with this new version of \FH. Numerically, we found that the terms 
beyond \order{\alt^2,\alt\alb} for which in previous versions of \FH the 
cancellation was incomplete are 
negligible for low scales ($\msusy \lesssim 0.5 \tev$). They can be more significant for high scales ($\sim 1.5 \gev$ for $\msusy \sim 20 \tev$). 

Furthermore, we investigated the impact of different parametrizations of
the non-logarithmic one- and two-loop terms. In this context, we found
the top-quark quark mass and the vev to be especially relevant. Despite the results
being formally identical at the strict two-loop level, using e.g.\ a
SM NNLO $\MS$ top quark mass instead of the OS top quark mass
induces changes in the higher-order non-logarithmic contributions. 

In our numerical comparison, we focused on a 
single scale scenario with a moderate $\tb$, which is particularly
suited for an EFT calculation.
We specifically compared the results of \FH and the EFT code \SHD. Using the NNLO value of the $\MS$ top Yukawa coupling in \SHD (by default the NNNLO value 
is used in \SHD, which leads to a downward shift by $\sim 0.5\gev$ in $M_h$),
we find very good agreement between the new version of \FH and \SHD
for scales $\msusy\gtrsim 1 \tev$. 
Such a good agreement is in fact expected for high SUSY scales since the
hybrid approach of \FH incorporates essentially the same logarithmic
contributions as pure EFT calculations. For $\msusy\lesssim 1 \tev$ we
observe significant differences between
\FH and \SHD due to terms suppressed by the SUSY scale that are not
incorporated in the EFT calculation of \SHD. The observed differences stay
relatively small for the considered simple scenario with a single SUSY
scale, reaching $\sim 1 \gev$ for $\msusy\sim 300 \gev$.
Larger deviations can be expected in SUSY scenarios with non-negligible mass
splittings between the various SUSY particles. Such kind of mass patterns
are accounted for in the diagrammatic fixed-order part of the hybrid approach.

The new version of \FH described in this paper, comprising an improvement in
the determination of the propagator poles and an option for 
using the $\DR$ scheme for the renormalization of the stop sector, will be
made public soon. 

The results obtained in this paper provide important input for an
improved estimate of the remaining theoretical uncertainties from unknown
higher-order corrections. 
In this context, we would like to stress once more that for the numerical
evaluations in this paper we have used a rather simple scenario where all SUSY
masses have been set to be equal to each other.
Having reconciled the hybrid approach of \FH with 
pure EFT calculations for this 
simple single scale scenario, we are now in a
position to assess the accuracy of the theoretical predictions also for 
more general scenarios with different hierarchies of scales. 
This will be analysed in a fothcoming publication.


\section*{Acknowledgments}
\sloppy{
We thank Pietro Slavich for useful discussions. H.B.\ is thankful to Thomas Hahn for his invaluable help
concerning all issues related to \FH.
H.B.\ and W.H.\ gratefully acknowledge support by the Deutsche Forschungsgemeinschaft (DFG) under Grant No.\ EXC-153 
(Excellence Cluster ``Structure and Origin of the Universe'').
The work of S.H.\ is supported in part by CICYT (Grant FPA 2013-40715-P),
in part by the MEINCOP Spain under contract FPA2016-78022-P,
in part by the ``Spanish Agencia Estatal de Investigacin'' (AEI) and the EU
``Fondo Europeo de Desarrollo Regional'' (FEDER) through the project
FPA2016-78645-P,
and by the Spanish MICINN's Consolider-Ingenio 2010 Program under Grant
MultiDark CSD2009-00064. The work  of  G.W.\  is  supported  in  part  by
the  DFG  through  the  SFB  676  ``Particles,  Strings and the Early
Universe'' and by the European Commission through the ``HiggsTools'' Initial
Training  Network  PITN-GA-2012-316704.
}


\appendix

\section{Fixed-order conversion: additional two-loop terms}\label{TLaddedterms}

In the limit $\msusy\gg M_t$ and degenerate $M_L=M_{\tilde t_R}=\msusy$, the one-loop contributions from the stop/top sector to the neutral Higgs self-energies at \order{\alt} are given by (for the remainder of this
section we drop the subscript ``$G_F$'', i.e.\ we use the shorthand 
$v \equiv v_{G_F}$)
\begin{align}
\hat\Sigma_{11} &= \frac{1}{16\pi^2} \frac{1}{\sbb}\frac{m_t^4}{v^2}\frac{\mu^2 \Xt^2}{M_S^4}, \\
\hat\Sigma_{12} &= \frac{1}{16\pi^2} \frac{1}{\sbb}\frac{m_t^4}{v^2}\frac{\mu \Xt}{M_S^2}\left[6-\frac{\Xt^2}{M_S^2}-\frac{1}{\tbe}\frac{\mu \Xt}{M_S^2}\right], \\
\hat\Sigma_{22} &= \frac{1}{16\pi^2} \frac{1}{\sbb}\frac{m_t^4}{v^2}\left[-12\ln\frac{M_S^2}{m_t^2}-12\frac{\Xt^2}{M_S^2}+\frac{\Xt^4}{M_S^4}-\frac{2}{\tbe}\frac{\mu \Xt}{M_S^2}\left(6-\frac{\Xt^2}{M_S^2}\right)+\frac{1}{\tbe^2}\frac{\mu^2 \Xt^2}{M_S^4}\right],
\end{align}
where $M_S^2 = \Mstope \Mstopz$, and $m_t$ is either the OS top mass or the $\MS$ SM top mass. We furthermore introduced the abbreviations
\begin{align}
s_x\equiv \sin x, \hspace{.5cm} c_x\equiv \cos x, \hspace{.5cm} t_x \equiv
\tan x .
\end{align}

If we convert the stop masses and the stop mixing parameter from the OS to the $\DR$ scheme using the shifts defined in \Eqss{XtConvEq}{MSt2ConvEq}, the following two-loop terms are generated (see \Eq{FOconvTLterm}),
\begin{align}
\Delta\hat\Sigma_{11} =& \frac{1}{8\pi^2} \frac{1}{\sbb}\frac{m_t^4}{v^2}\left[\frac{\Delta \Xt}{M_S}\frac{\mu^2 \Xt}{M_S^3}-2\frac{\Delta M_S}{M_S}\frac{\mu^2 \Xt^2}{M_S^4}\right], \\
\Delta\hat\Sigma_{12} =& \frac{1}{16\pi^2} \frac{1}{\sbb}\frac{m_t^4}{v^2}\left[\frac{\Delta \Xt}{M_S}\left(-3\frac{\mu \Xt^3}{M_S^3}-\frac{2}{\tbe}\frac{\mu^2 \Xt}{M_S^3}+6\frac{\mu}{M_S}\right)+\frac{\Delta M_S}{M_S}\left(4\frac{\mu \Xt^3}{M_S^4}+\frac{4}{\tbe}\frac{\mu^2 \Xt^2}{M_S^4}-12\frac{\mu \Xt}{M_S^2}\right)\right],\\
\Delta\hat\Sigma_{22} =& \frac{1}{8\pi^2} \frac{1}{\sbb}\frac{m_t^4}{v^2}\left[\frac{\Delta \Xt}{M_S}\left(-2\frac{\Xt}{M_S}\left(6-\frac{\Xt^2}{M_S^2}\right)-\frac{3}{\tbe}\frac{\mu}{M_S}\left(2-\frac{\Xt^2}{M_S^2}\right)+\frac{1}{\tbe^2}\frac{\mu^2 \Xt}{M_S^3}\right) \right.\nonumber\\
&\left.\hspace{1.84cm}-2\frac{\Delta M_S}{M_S}\left(6-6\frac{\Xt^2}{M_S^2}+\frac{\Xt^4}{M_S^4}-\frac{2}{\tbe}\frac{\mu \Xt}{M_S^2}\left(3-\frac{\Xt^2}{M_S^2}\right)+\frac{1}{\tbe^2}\frac{\mu^2 \Xt^2}{M_S^4}\right)\right].
\end{align}
The quantity $\Delta M_S$ is given by
\begin{align}
\Delta M_S = \frac{1}{2}\left(\frac{\Delta \mstope}{\Mstope}+\frac{\Delta
  \mstopz}{\Mstopz}\right)M_S, 
\end{align}
where $\Delta \Xt$ and $\Delta m_{\tilde t_{1,2}}$ are defined in \Eqss{XtConvEq}{MSt2ConvEq}.

Note that for all numerical results presented in this work, we used expressions valid also for low $\msusy$ ($\msusy\sim M_t$) and general SUSY breaking. Note also that the shifts are performed for all self-energies and not only for the $hh$ self-energy as shown exemplary in \Sec{FHwithDRinputSection}. Therefore, the procedure remains also valid in non-decoupling scenarios ($M_A \sim M_Z$).

As described in \Sec{FHwithDRinputSection}, these two-loop terms are finally
added to the respective self-energies, i.e., the $\Delta\hat\Sigma$'s are
added to the two-loop self-energies obtained from the diagrammatic
calculation. Higher-order
terms which would be generated by a scheme conversion of the input
parameters are omitted. In this way, the renormalization of the stop sector
is changed from the OS to the $\DR$ scheme. This alternative renormalization
scheme will be available as an option in the next \FH version.


\section{Logarithms arising from the determination of the propagator pole}\label{p2TermsApp}

In this Appendix, we give explicit expressions,
valid in the decoupling limit, 
for the logarithms induced by
the momentum dependence of the non-SM contributions to the MSSM Higgs
self-energy, i.e.\ for the quantity $\Delta_{p^2}^\text{logs}$ 
defined in \Eq{p2TermsLog}.

In order to derive the $(n+1)$th order iterative solution to the Higgs
pole mass equation (see \Eq{HiggsPoleEq}) in terms of lower order solutions,
F\`{a}a di Bruno's formula (extended chain rule for derivatives) is used,
\begin{align}\label{RecursionFormula}
&(M_h^2)^{(n+1)}=- \sum_{(a_1,...,a_n)\in T_n} \frac{1}{a_1 !\cdot ...\cdot
a_n !}\cdot\left[\left(\frac{\partial}{\partial
p^2}\right)^{(a_1+...+a_n)}\hat\Sigma_{hh}^\MSSM(p^2)\right]_{p^2=m_h^2}\cdot
\prod_{m=1}^n (M_h^2)^{(m)} ,
\end{align}
where an $n$-tuple of non negative integers $(a_1,...,a_n)$ is an element of $T_n$ if $1\cdot a_1+2\cdot a_2+...+n\cdot a_n = n$. 

The zeroth order correction
\begin{align}
(M_h^2)^{(0)} = m_h^2 
\end{align}
serves as starting point of the recursion.

We split $\Delta_{p^2}^{\text{logs}}$ into a leading, a next-to-leading
and a next-to-next-to-leading logarithm piece, 
\begin{align}
\Delta_{p^2}^\text{logs} = \Delta_{p^2}^{\text{LL}} +
\Delta_{p^2}^{\text{NLL}} + \Delta_{p^2}^{\text{NNLL}} + \ldots \; .
\end{align}
In \FH, the full momentum dependence by default is taken into
account only at the one-loop level. At the two-loop level, the external
momentum is set to zero (see \cite{Borowka:2014wla,Borowka:2015ura} for
a discussion of the momentum dependence at the two-loop level). We can
therefore
split up the non-SM contributions to the Higgs self-energy into a one- and a
two-loop piece, 
\begin{align}
\hat\Sigma_{hh}^\nonSM(p^2) = \hat\Sigma_{hh}^{\nonSM,(1)}(p^2) + \hat\Sigma_{hh}^{\nonSM,(2)}(0). 
\end{align}

To shorten the expressions for the individual contributions, we first introduce abbreviations. We write the non-SM contributions to the Higgs self-energy as
\begin{align}
\hat\Sigma_{hh}^{\nonSM,(1)}(m_h^2) &= k\left(c_{1,1}^\chi L_\chi + c_{1,1}^A L_A+ c_{1,1}^{\tilde f} L_S + c_{1,0}\right), \\
\hat\Sigma_{hh}^{\nonSM,(2)}(0) &= k^2\left(c_{2,2} L_S^2 + c_{2,1} L_S + c_{2,0}\right), 
\end{align}
where $k\equiv(4\pi)^{-2}$ is used to keep track of the loop order and
\begin{align}
L_\chi\equiv\ln\frac{M_\chi^2}{m_t^2}, \hspace{.5cm} L_A\equiv\ln\frac{\MA^2}{m_t^2}, \hspace{.5cm} L_S\equiv\ln\frac{\msusy^2}{m_t^2}.
\end{align}
Here it should be noted that in this work we set 
\begin{align}
M_\chi\equiv M_1 = M_2 = \mu 
\quad \mbox{and} \quad M_\chi = \MA = \msusy\,.
\end{align}
In this Appendix, however, we keep them independent to be able to use
the expressions also for more general cases. 

The subscript of a coefficient $c_{a,b}$ indicates that it is the prefactor of the term $k^a L^b$ ($L=L_\chi,L_A,L_S$). The corresponding superscript marks the origin of the respective term (from EWinos $\chi$, from heavy Higgses $A$ or from sfermions $\tilde f$). These superscripts are used only at the one-loop level to be able to differentiate between the different types of appearing logarithms ($L_\chi$, $L_A$ and $L_S$). In the $\DR$ scheme, the appearing coefficients up to \order{v^2/M_{\text{heavy}}^2} ($M_{\text{heavy}}=M_\chi,\MA,\msusy$) are given by (for the remainder of this
section we drop the subscript ``$G_F$'', i.e.\ we use the shorthand 
$v \equiv v_{G_F}$)
\begin{align}
c_{1,1}^{\tilde f} &= -2v^2\left[6 y_t^4+\frac{3}{2}y_t^2(g^2+g'^2)c_{2\beta}+\frac{1}{2}g^4+\frac{5}{6}g'^4+\frac{1}{6}(3g^4 + 5 g'^4)c_{4\beta}\right], \\
c_{1,1}^{\chi} &= -2v^2\left[\frac{1}{24}g'^4(-11+c_{4\beta})-\frac{3}{8}g^4(5+c_{4\beta})-g^2 g'^2 s_{2\beta}^2\right],\\
c_{1,1}^{A} &= -2v^2\left[\frac{1}{192}g^4(53-28c_{4\beta}-9c_{8\beta})+\frac{1}{192}g'^4(29-4c_{4\beta}-9c_{8\beta})+\frac{1}{8}g^2 g'^2(5+3c_{4\beta})s_{2\beta}^2\right], \\
c_{1,0} &= -2v^2\Bigg\{ 6 y_t^2 \left[\left(y_t^2+\frac{1}{8}(g^2+g'^2)c_{2\beta}\right)\xf^2-\frac{1}{12}y_t^2\xf^4\right] \nonumber\\
&\hspace{1.4cm}- \frac{1}{4}y_t^2(g^2+g'^2)\xf^2 c_{2\beta}^2 -\frac{3}{16}(g^2+g'^2)^2 s_{4\beta}^2 \nonumber\\
&\hspace{1.4cm}-\left[\left(\frac{3}{4}-\frac{1}{6}c_{2\beta}^2\right)g^4+\frac{1}{2}g^2g'^2+\frac{1}{4}g'^4\right]\nonumber\\
&\hspace{1.4cm}+\frac{1}{24}(\sbe+\cbe)^2\cdot\nonumber\\
&\hspace{1.8cm}\cdot\bigg[-51g^4-24g^2g'^2-13g'^4 +\left((g^2+g'^2)c_{4\beta}+2(g^2-g'^2)s_{2\beta}\right)(3g^2+g'^2)\bigg]\Bigg\},\\
c_{2,2} &= - 2 v^2 y_t^4\left(-48g_3^2+9y_t^2\right), \\
c_{2,1} &= - 2 v^2 y_t^4\Big[8 g_3^2 \left(4-12\xf^2+\xf^4\right)-\frac{3}{2}y_t^2\left(20-12\xf^2+\xf^4\right)\Big],
\end{align}
where all appearing couplings are SM $\MS$ couplings evaluated at
$Q=M_t$ ($g$, $g'$ are the electroweak gauge couplings, and $\xf\equiv
\Xt/\msusy$). We write the derivative of the non-SM contributions to the Higgs self-energy as
\begin{align}
\hat\Sigma_{hh}^{\nonSM,(1)\prime}(m_h^2) &= k\left(c'_{1,1} L_\chi +
c'_{1,0}\right) ,
\end{align}
with the primes denoting that the corresponding coefficient appears in the derivative of the self-energy. We again drop contributions of \order{v^2/M_{\text{heavy}}^2}. The coefficient multiplying $L_\chi$ originates purely from EWino graphs and reads
\begin{align}
c'_{1,1} = -\frac{1}{2}(3g^2+g'^2).
\end{align}
The non-logarithmic coefficient has contributions from EWinos as well as from
stops (neglecting all other Yukawa couplings), 
\begin{align}\label{dsehh_stop_Eq}
c'_{1,0} = \underbrace{\frac{1}{2}y_t^2\xf^2}_{\text{stop contr.}}\underbrace{-\frac{1}{6}(3g^2+g'^2)(\sbe+\cbe)^2}_{\text{EWino contr.}}.
\end{align}
All higher derivatives of $\hat\Sigma_{hh}^\nonSM(p^2)$ are suppressed, i.e.\ of \order{p^2/M_{\text{heavy}}^2}.

The SM contributions are written in a similar way,
\begin{align}
\left.\left(\frac{\partial}{\partial
p^2}\right)^n\hat\Sigma_{hh}^{\SM,(1)}(p^2)\right|_{p^2=m_h^2} &= k\tilde
c^{n}_1 ,
\end{align}
where the superscript '$n$' denotes the $n$th derivative of $\hat\Sigma_{hh}^{\SM,(1)}$. Here, we only give explicit expressions for the pure top Yukawa contributions to the first five derivatives of $\hat\Sigma_{hh}^{\SM,(1)}$,
\begin{align}
\tilde c_1^{(1)} &= - \frac{1}{2} y_t^2 v^0,\label{dSEtop1}\\
\tilde c_1^{(2)} &= \frac{3}{5} y_t^0 v^{-2}, \\
\tilde c_1^{(3)} &= \frac{9}{70} y_t^{-2} v^{-4}, \\
\tilde c_1^{(4)} &= \frac{2}{35} y_t^{-4} v^{-6}, \\
\tilde c_1^{(5)} &= \frac{4}{77} y_t^{-6} v^{-8}. \label{dSEtop5}
\end{align}

Eq.~(\ref{RecursionFormula}) allows now to successively derive all corrections induced by the momentum dependence of the non-SM contributions to the $hh$ self-energy. The generated leading logarithms can be resummed easily, since higher derivatives of $\hat\Sigma_{hh}^\nonSM$ are always suppressed, as noted before. The resummed expression is given in terms of the $c$ coefficients by
\begin{align}
\Delta_{p^2}^{\text{LL}} = k^2 \frac{c'_{1,1} L_\chi}{1 + k c'_{1,1} L_\chi} \left[c_{1,1}^\chi L_\chi + c_{1,1}^A L_A + c_{1,1}^{\tilde f} L_S + k c_{2,2} L_S^2\right].
\end{align}
A similar expression can be derived at the NLL level. We obtain 
\begin{align}
\Delta_{p^2}^{\text{NLL}} =& k^2 \frac{1}{(1 + k c'_{1,1} L_\chi)^2}\cdot\nonumber\\
&\cdot\Big[c_{1,1}^\chi c'_{1,0} L_\chi + c_{1,1}^A c'_{1,0} L_A + c_{1,1}^{\tilde f} c'_{1,0} L_S + c_{1,0} c'_{1,1} L_\chi \nonumber\\
&\hspace{.3cm}+ k\left( c_{1,0} (c'_{1,1})^2 L_\chi^2 + c_{2,1}c_{1,1}' L_\chi L_S + c_{2,2}c_{1,0}' L_S^2\right)\nonumber\\
&\hspace{.3cm}+k^2 c_{2,1} (c_{1,1}')^2 L_\chi^2 L_S\Big].
\end{align}
At the NLL level however, additional terms proportional to derivatives
of the light self-energy exist. Since these derivatives are not
suppressed by a heavy mass, it seems not to be possible to resum the
corresponding logarithms. Nevertheless, including terms up to the 7-loop
order we find a good convergence behavior and an induced shift of 
\order{\pm 2 \gev^2} to $M_h^2$
in the parameter region $M_t < M_{\text{heavy}}
\lesssim 20 \tev$. The respective shift in $M_h$ is of \order{50 \mev}. We
therefore neglect this contribution completely. 

At the NNLL level, we take into account only terms proportional to the strong
gauge coupling and the top-Yukawa coupling (terms proportional to electroweak
gauge couplings are negligible). We find that at this level all terms include
derivatives of the SM self-energy. We also find that this contribution to
$M_h^2$ is not
negligible, \order{20 \gev^2}. Therefore, we include terms up to the
7-loop order, which are given by 
\begin{align}
 \Delta_{p^2}^{\text{NNLL}} =& k^3 L_S c'_{1,0}\left[c_{2,1} - c_{1,1}^{\tilde f} \tilde c_{1}'\right] \nonumber\\
& - k^4 L_S^2 c'_{1,0} \left[c_{2,2}c'_{1,0}+c_{2,2} \tilde c_{1}^{(1)}-\frac{1}{2}\left(c_{1,1}^{\tilde f}\right)^2\tilde c_{1}^{(1)}\right]\nonumber\\
& + k^5 L_S^3 c'_{1,0}\left[c_{1,1}^{\tilde f}c_{2,2}\tilde c_{1}^{(2)}-\frac{1}{6}\left(c_{1,1}^{\tilde f}\right)^3 \tilde c_{1}^{(3)}\right] \nonumber\\
& + \frac{1}{2}k^6 L_S^4 c'_{1,0}\left[\left(c_{2,2}\right)^2 \tilde c_{1}^{(2)}-c_{2,2}\left(c_{1,1}^{\tilde f}\right)^2 \tilde c_{1}^{(3)}+\frac{1}{12}\left(c_{1,1}^{\tilde f}\right)^4\tilde c_{1}^{(4)}\right] \nonumber\\
& -\frac{1}{2}k^7 L_S^5 c'_{1,0}\left[\left(c_{2,2}\right)^2 c_{1,1}^{\tilde f} \tilde c_{1}^{(3)}-\frac{1}{3}c_{2,2}\left(c_{1,1}^{\tilde f}\right)^3 \tilde c_{1}^{(4)}+\frac{1}{60}\left(c_{1,1}^{\tilde f}\right)^5\tilde c_{1}^{(5)}\right]\nonumber\\
& + \order{k^8},
\end{align}
where all terms in the $c$ coefficients proportional to $g$ or $g'$ are set to zero. Correspondingly, the derivatives of the light self-energy only include terms proportional to $y_t$. These are listed in \hbox{Eqs. (\ref{dSEtop1})-(\ref{dSEtop5})}.  This loop expansion  quickly converges such that we can safely drop higher-order contributions (8-loop and beyond). 

We find the electroweak contributions at the NNLL level and even 
higher-order logarithms (N$^n$L with $n>2$) to be completely negligible.
Similar expressions can
easily be obtained for the non-logarithmic terms of the same origin (see
\Eq{p2NonLogDef}).


\newpage

\bibliographystyle{JHEP}
\bibliography{bibliography}{}

\end{document}